\documentclass[runningheads]{llncs}

\usepackage{eccv}

\usepackage{eccvabbrv}

\usepackage[table]{xcolor}

\usepackage{amsmath}
\usepackage{amsbsy}
\usepackage{amssymb}
\usepackage{amsfonts}

\usepackage{amsthm}
\usepackage{physics}
\usepackage{datetime}
\usepackage{graphicx}
\usepackage{algorithmic}
\usepackage{algorithm}
\usepackage{enumerate}

\providecommand{\cref}[1]{Chapter~\ref{#1}}

\providecommand{\fref}[1]{Figure~\ref{#1}}

\providecommand{\R}{\ensuremath{\mathbb{R}}}

\providecommand{\C}{\ensuremath{\mathbb{C}}}
\providecommand{\E}{\ensuremath{\mathbb{E}}}

\providecommand{\abs}[1]{\lvert#1\rvert}
\providecommand{\norm}[1]{\lVert#1\rVert}

\renewcommand{\vec}[1]{\ensuremath{\mathbf{#1}}}
\providecommand{\greekvec}[1]{\ensuremath{\boldsymbol{#1}}}
\providecommand{\mat}[1]{\ensuremath{\mathbf{#1}}}

\providecommand{\calN}{\mathcal{N}}

\providecommand{\calP}{\mathcal{P}}

\providecommand{\mA}{\mat{A}}

\providecommand{\mC}{\mat{C}}

\providecommand{\mF}{\mat{F}}

\providecommand{\mI}{\mat{I}}

\providecommand{\mP}{\mat{P}}

\renewcommand{\va}{\vec{a}}
\renewcommand{\vb}{\vec{b}}
\providecommand{\vc}{\vec{c}}
\providecommand{\vd}{\vec{d}}

\providecommand{\vn}{\vec{n}}

\providecommand{\vp}{\vec{p}}

\providecommand{\vx}{\vec{x}}
\providecommand{\vy}{\vec{y}}
\providecommand{\vz}{\vec{z}}

\providecommand{\vtheta}{\greekvec{\theta}}

\providecommand{\vphi}{\greekvec{\phi}}

\providecommand{\vxhat}{{\widehat{\vx}}}

\providecommand{\vzero}{\vec{0}}
\providecommand{\vone}{\vec{1}}

\newcommand{\argmin}[1]{\mathop{\underset{#1}{\mbox{argmin}}}}

\usepackage{graphicx}

\usepackage{booktabs}
\usepackage{float}

\usepackage[accsupp]{axessibility}  

\usepackage[pagebackref,breaklinks,colorlinks,citecolor=eccvblue]{hyperref}
\usepackage{hyperref}

\usepackage{orcidlink}

\newcommand{\rev}[1]{\textcolor{black}{#1}} 

\begin{document}

\title{Pupil Design for Computational Wavefront Estimation} 

\author{Ali Almuallem\inst{1}\orcidlink{0000-0002-5235-077X} \and
Nicholas Chimitt\inst{1}\orcidlink{0000-0001-9528-0102} \and
Bole Ma\inst{1}\orcidlink{0009-0007-0417-3979} \and
Qi Guo\inst{3}\orcidlink{0000-0002-8329-7668} \and
Stanley H. Chan\inst{3}\orcidlink{0000-0001-5876-2073}}

\authorrunning{A. Almuallem et al.}

\institute{Elmore Family School of Electrical and Computer Engineering, Purdue University, USA}

\maketitle

\begin{abstract}

    Establishing a precise connection between imaged intensity and the incident wavefront is essential for emerging applications in adaptive optics, holography, computational microscopy, and non-line-of-sight imaging. While prior work has shown that breaking symmetries in pupil design enables wavefront recovery from a single intensity measurement, there is little guidance on how to design a pupil that improves wavefront estimation. In this work we introduce a quantitative asymmetry metric to bridge this gap and, through an extensive empirical study and supporting analysis, demonstrate that increasing asymmetry enhances wavefront recoverability. We analyze the trade-offs in pupil design, and the impact on light throughput along with performance in noise. Both large-scale simulations and optical bench experiments are carried out to support our findings.

  \keywords{Wavefront estimation \and Computational imaging \and Asymmetry}

\end{abstract}

\section{Introduction}
\label{sec:intro}
As computer vision increasingly integrates with wave-optics-based systems, a precise connection between imaged intensity and the incident wavefront becomes necessary. Many modern applications, such as holographic displays \cite{hossein2020deepcgh, chakravarthula2020learned}, computational microscopy \cite{zhao_2025_fluo_microscopy}, adaptive optics \cite{jiang2026guidestar}, ptychographic methods \cite{zheng2013wide, tian20153d}, and non-line-of-sight imaging \cite{lindell2019wave, o2018confocal} rely on image formation models that incorporate the wave behavior of light. However, conventional sensors only capture intensity, losing critical phase information. Recovering the lost phase information is fundamental to the success of computer vision systems in these domains that go beyond purely intensity-based models \cite{shechtman_2015_a}.

The central challenge in this domain is the recovery of a complex wavefront from its Fourier magnitude, known as Fourier phase retrieval \cite{Fienup_1982_Comparison, Unser_2023_Review}. Historically, solutions to this ill-posed problem include strong priors for specialized domains, multiple measurements by optical methods \cite{gonsalves_1982_a, Feng_2023_NeuWS}, or specialized optical hardware such as lenslet arrays \cite{Tyson_Frazier_Book}. The fundamental bottleneck for machine learning \rev{remains the} lack of uniqueness given a single intensity measurement. As a result, any learning-based solution would require a similarly complicated optical design or modeling, making the problem highly specialized.

\begin{figure}[h]
    \centering
    \includegraphics[width=0.98\linewidth]{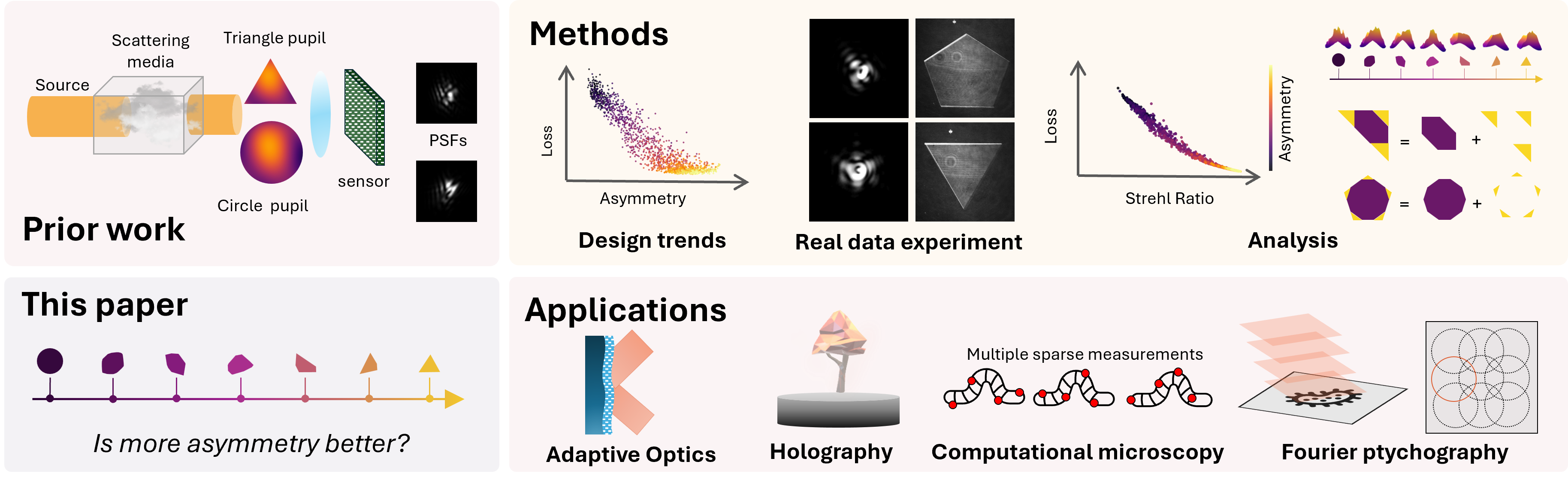}
    \caption{Our work investigates the role of pupil design in wavefront estimation, proposes an asymmetry metric to gauge pupil performance, and provides thorough empirical evidence to support our findings. It directly impacts fields like adaptive optics and microscopy, where accurate wavefront estimation is crucial.}
    \label{fig: fig1}
\end{figure}

Recently, it was shown that machine learning can recover a wavefront from a single intensity measurement by symmetry breaking in camera pupil design \cite{Chimitt_Almuallem}. This is possible for problems related to adaptive optics, microscopy, and other imaging problems where the pupil may be modified to be asymmetric, similar to coded apertures that have been used in numerous computer vision applications \cite{Levin_2007_CodedAperture, Candes_2015_Coded, asif_2016_a}. However, despite this theoretical guarantee, it does not offer guidance on \textit{how} to design the pupil. In this work, we provide an extensive empirical study aimed at addressing the trade-off in pupil design for wavefront recovery and analyze the performance in the presence of noise.

We illustrate the context of this paper in \fref{fig: fig1} and summarize our contributions as follows: 
\begin{enumerate}
    \item \textbf{Problem definition.} We define the pupil design problem and introduce an asymmetry metric to characterize a pupil's capacity to resolve ambiguities. This metric serves as the basis of both analytic and empirical results.
    \item \textbf{Empirical study of design.} We study the impact of asymmetry by utilizing a dataset of 7500 pupils and 6000 phase aberrations, along with two baseline networks to understand the gap in training and testing performance across network architectures.
    \item \textbf{Real experiments.} We offer prototype-level real experiments that support the findings of our simulations.
\end{enumerate}

\section{Related Work}
Phase retrieval is concerned with the recovery of a vector $\vx \in \C^N$ from \rev{a measurement} $\abs{\mA \vx}^2$. When $\mA \in \C^{N \times N}$ is the discrete Fourier transform matrix, the problem is Fourier phase retrieval \cite{Fienup_1982_Comparison, fienup1978reconstruction, jaganathan2016phase}. Although oversampling can help alleviate issues in uniqueness \cite{Hayes_1982_Multidimensional, Hayes_1982_Reducible}, there remain ambiguities that can not be eliminated by sampling alone. Wavefront estimation requires eliminating some of these ambiguities \cite{Chimitt_Almuallem}. Because of the similarity of these two problems, we discuss them both to comment on their domains of application and differences.

\subsubsection{Phase retrieval} Phase retrieval has been studied for many decades \cite{sayre1952some, gerchberg_1972_a, Fienup_1982_Comparison, Hayes_1982_Multidimensional, jaganathan2016phase, Unser_2023_Review} and arises in a number of problems such as crystallography \cite{millane_1990_a, Sayre_2002_Xray}, holography \cite{hossein2020deepcgh}, and Fourier ptychography \cite{zheng2013wide, tian20153d}. Long-standing non-convex algorithms such as the Gerchberg-Saxton \cite{gerchberg_1972_a} and Hybrid Input-Output \cite{Fienup_1982_Comparison} are still popular today, while alternatives \cite{Candes_2015_Wirtinger, YueLu_2019_SpectralInit, ranieri_2013_a} are designed for random coded measurements. 

More recently, phase retrieval algorithms have been \rev{developed} using machine learning techniques. These range from iterative techniques \cite{Metzler_2018_prDeep}, better initializations \cite{paine2018machine}, and having an added reference signal to overcome ambiguous solutions \cite{hyder2020solving}. Dataset preprocessing has also been shown to help overcome issues related to non-uniqueness \cite{JuSun_2021_BreakSymmetry, zhang2024s}. For generic phase retrieval problems and machine learning solutions, we refer the interested reader to Wang et al. \cite{EdmundLam_2024_Review} for a thorough review.

\subsubsection{Wavefront estimation} Wavefront estimation is critical for adaptive optics for imaging through scattering \cite{Tyson_Frazier_Book, Chan_TurbulenceBook}, calibration \cite{fienup1993hubble, Fienup_1993_Complicated}, and microscopy with many methods using multiple diverse measurements \cite{gonsalves_1979_a, gonsalves_1982_a}. More recently, neural representations have been shown to be a viable alternative, enabling new capabilities such as passive wavefront estimation and correction \cite{Feng_2023_NeuWS, xie_2024_a}. All methods described here require multiple measurements.

For real-time applications such as adaptive optics, hardware solutions are often utilized \cite{Tyson_Frazier_Book}. Examples of hardware solutions range from Shack-Hartmann sensors, pyramid wavefront sensors, and interferometric methods \cite{Wyant_2003_Interferometry}. While oftentimes the optical hardware increases in complexity, the measurements avoid the need to solve the Fourier phase retrieval problem.

Recent single measurement methods have high potential to impact real-time applications and inverse problems. Pre-conditioning optics \cite{nishizaki_2019_a} is one example, though asymmetry in support pupils has long been known to improve recovery \cite{cederquist_1989_a, Martinache_2013_Asymmetric}. Recently, it was shown to enable a unique solution for the wavefront estimation problem \cite{Chimitt_Almuallem, chimitt2024phase}, though there is little understanding of how to define asymmetry for this problem or how best to design an asymmetric pupil.

\section{Pupil Design for Wavefront Estimation}
This section is aimed at defining the pupil design problem, how we measure the amount of asymmetry present in a pupil, and providing analytic results that support our later empirical findings.

\subsection{Problem definition}
\label{sec: probscope}
\begin{figure}
    \centering
    \includegraphics[width=0.95\linewidth]{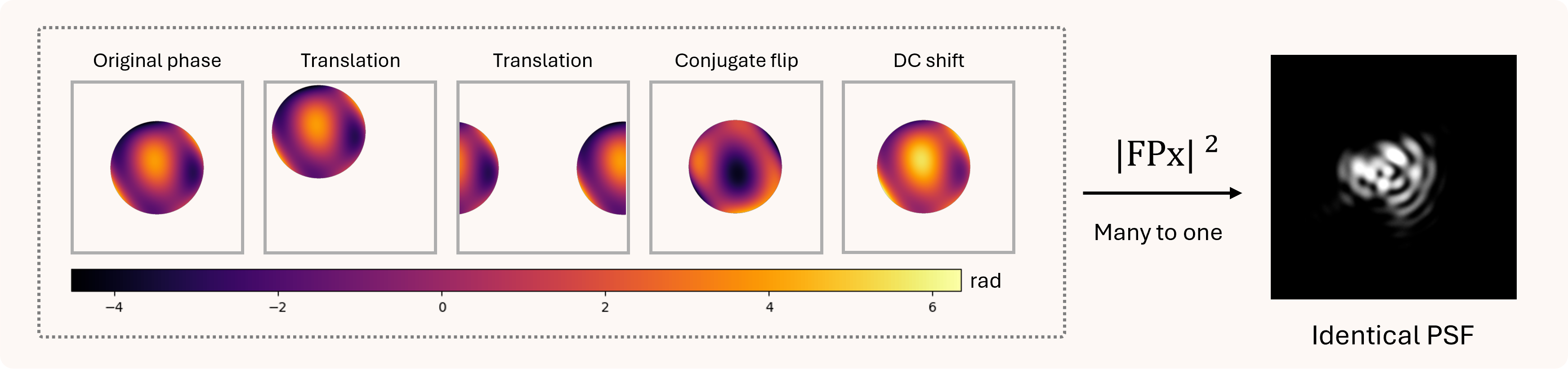}
    \caption{Visualization of trivial ambiguities. Different phase aberrations (or any combinations of them) yield the same point spread function (PSF), resulting in a \rev{many-to-one} relationship for the wavefront estimation problem.}
    \label{fig: trivial_ambiguities}
\end{figure}

The point spread function (PSF) of an imaging system with a pupil support $\mP \in \R^{N \times N}$ represented as a diagonal matrix and incident wave $\vx$ is defined as \cite{Goodman_FourierBook}
\begin{equation}
    \vy = \abs{\mF \mP \vx}^2 + \vn,
\end{equation}
where $\mF$ is the Fourier transform matrix,  $\vn \sim \calN(\vzero, \sigma^2 \mI)$ and $\abs{\cdot}$ denotes the elementwise absolute value of a vector. The recovery of $\mP \vx$ from $\vy$ is non-unique; although oversampling can eliminate many ambiguous solutions \cite{Hayes_1982_Multidimensional, Hayes_1982_Reducible}, there remain ambiguities consisting of the composition of translations, a unit magnitude complex multiple, and a conjugate flip \cite{jaganathan2016phase}. For wavefront estimation, it is necessary to eliminate translation and conjugate flip ambiguities, the latter we denote as $\vx_*$. We will assume $\vx$ to be a unit magnitude vector, i.e., $\abs{\vx} = \vone$, with phase
\begin{equation}
    \vphi = \sum_{m = 1}^M a_m \vz_m,
\end{equation}
hence $\vx = \exp(j \vphi)$ and $\vz_m$ is the $m$th basis vector, e.g., a Zernike polynomial \cite{Noll_1976}. We illustrate these ambiguous solutions in \fref{fig: trivial_ambiguities} and refer readers to \cite{Chimitt_Almuallem} for further detail. 

As a common application of wavefront estimation is adaptive optics, an important metric will be the ability of the recovered signal $\vxhat$ to serve as an optical correction, mathematically represented as $\abs{\mF \mP (\vx \odot \vxhat^*)}^2$. All pupils considered in this paper can be inscribed within a maximum circular support $\mC$, hence all pupils satisfy $\mC \mP = \mP$. We first define a Strehl ratio of the system with pupil $\mP$ relative to the unmodified system with pupil $\mC$
\begin{equation}
    \rho_{\mC}(\mP, \vx, \vxhat) = \frac{\max \abs{\mF \mP (\vx \odot \vxhat^*)}^2}{\max \abs{\mF \mC \vone}^2}.
    \label{eq: strehl}
\end{equation}
We will also define $\rho_{\mP}(\mP, \vx, \vxhat)$ to be the Strehl ratio of a pupil relative to itself \rev{(i.e., with $\max \abs{\mF \mP \vone}^2$ in the denominator)}. All the reported Strehl ratios are of the second type (relative to the pupil itself) unless otherwise specified. If $\vxhat = \vx$, the correction will be optimal and $\rho_{\mP} = 1$, but if $\vxhat = \vx_*$, i.e., the conjugate flip solution, it can often be worse than with no correction.

While maximizing the expectation of either Strehl ratios would serve as a valid objective function, in practice we find that it provides weak and inconsistent supervision as it only relies on maximum values. Instead, due to the uniqueness provided by symmetry breaking, a standard MSE loss can be utilized, which we do in a two-step fashion. We first find the minimum mean square error (MMSE) estimator for $\vx \sim p(\vx)$ and $\mP \sim p(\mP)$,
\begin{equation}
    \vtheta^* = \argmin{\vtheta} \; \E \norm{f_{\vtheta} \left(\vy, \mP \right) - \mP \vx}^2.
    \label{eq: loss_fn}
\end{equation}
By design of the distribution $p(\mP)$, we can avoid stagnation due to symmetries or trivial solutions, e.g., $\mP = \vzero$. The second step involves optimizing or combinatorially searching for the pupil that satisfies
\begin{equation}
    \mP^* = \argmin{\mP \in \calP} \; \E \norm{f_{\vtheta} \left(\vy, \mP \right) - \mP \vx}^2.
    \label{eq: pupil_opt}
\end{equation}
where $\calP$ is a feasible set of pupils, e.g., the set of convex hulls within the circular pupil $\mC$. The primary results of this paper are based on the performance of wavefront recovery across a set of sampled pupils. We train our network using the MSE loss \eqref{eq: loss_fn}, but further establish a connection between performance in MSE and Strehl ratio correction.

\subsection{Significance of ambiguities}
\label{sec: overfitting_reason}
An asymmetric pupil was motivated by the purpose of ``breaking the symmetry'' inherent to a symmetric pupil. Mathematically, for a solution $\mP \vx$, there exists another potential solution $(\mP \vx)_*$, where the underset $*$ denotes a complex conjugate and flip about the origin. Because $(\mP \vx)_* = \mP_* \vx_* = \mP \vx_*$, the two solutions $\vx$ and $\vx_*$ lie on the same support and are indistinguishable by their Fourier magnitude due to standard Fourier properties. We now discuss the impact of the conjugate flip ambiguity on the MMSE estimator and the cost of recovering this solution as it relates to wavefront estimation.

Suppose that $\mP$ is not symmetric, i.e., $\mP_* \neq \mP$, but $\abs{\mF \mP \vx}^2 \approx \abs{\mF \mP \vx_*}^2$. This can occur regularly in practice when there is a significant overlap \rev{between} $\mP$ and $\mP_*$. Due to the similarity in the PSFs, the two may easily become ambiguous in the presence of noise and lead to the erroneous recovery of $\vx_*$. This can be understood by its impact on the MMSE estimator,
\begin{equation}
    \int \vx \cdot p(\vx \vert \vy) d\vx \approx \pi_1 \mP \vx + \pi_2 \mP \vx_*,
\end{equation}
where we have approximated $p(\vx \vert \vy) \approx \pi_1 \calN(\mP \vx, \Sigma_1) + \pi_2 \calN(\mP \vx_*, \Sigma_2)$. The bi-modal-like behavior has a significant impact on the training behavior of a network. Suppose that the MMSE estimator provides a weighted combination of $\vxhat = \pi_1 \vx + \pi_2 \vx_*$. If $\pi_1 \approx \pi_2$, $\abs{\mF \mP \vxhat}^2$ will differ significantly from the observed PSF due to the non-linearity of the forward model. Accordingly, the penalty for such a decision will be large when training the network. This drives the network to overfit to the data, consistent with behavior described in prior work \cite{Chimitt_Almuallem, JuSun_2021_BreakSymmetry}.

\subsection{A definition of asymmetry for wavefront estimation}
\begin{figure}[H]
    \centering
    \includegraphics[width=0.85\linewidth]{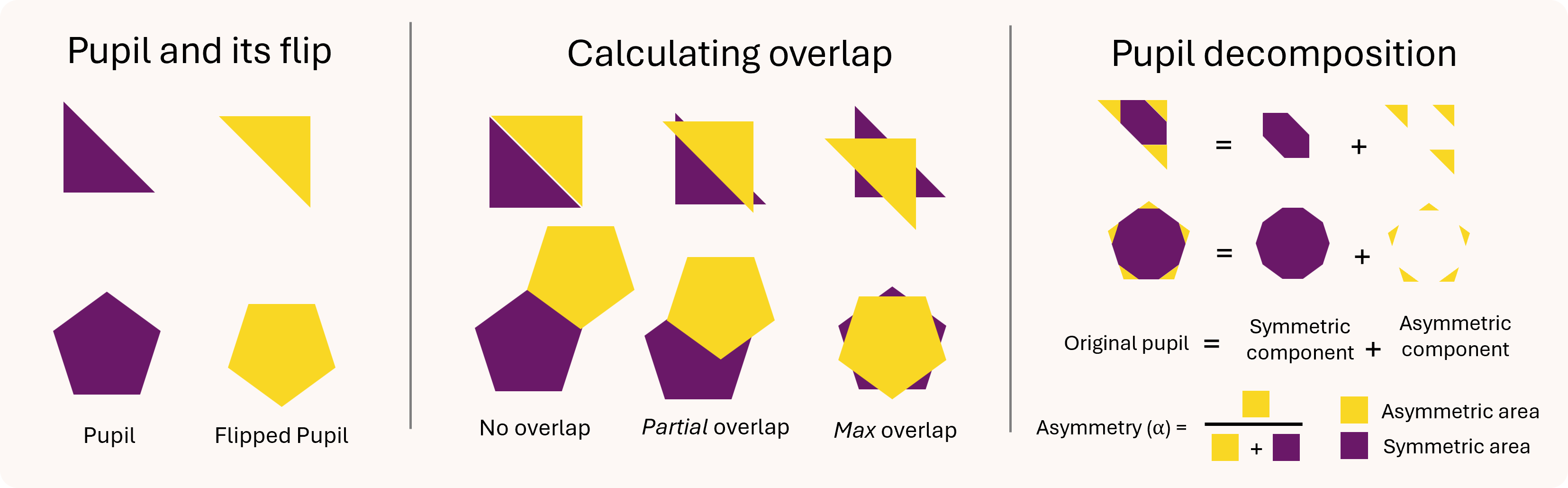}
    \caption{The asymmetry metric $\alpha$ is defined as the maximum non-overlapping area between the pupil and its flip about its center. A pupil can therefore be decomposed into two parts: a symmetric part that is invariant over flipping, and an asymmetric part. The symmetric part of the pupil contributes to the ambiguities in the intensity measurement, while the asymmetric parts encode distinguishable intensities.}
    \label{fig:asymmetry_levels}
\end{figure}

To quantify how asymmetric one pupil is versus another, we introduce a metric to measure the asymmetry of a pupil. Due to the nature of trivial ambiguities, our definition of asymmetry includes the potential for shifts, i.e., translations of the recovered phase and pupil. We define the asymmetry value to be
\begin{equation}
    \alpha = 1 - \frac{\max (\vp \circledast \vp)}{\norm{\vp}},
\end{equation}
where $\circledast$ denotes a 2D convolution and $\vp = \text{diag}(\mP)$. Interpreting this expression, we first compute the maximum amount of overlap a pupil has with any translation of its flipped version. We then normalize by the total pupil area. The result is a \textit{relative} measure of how much the pupil maximally overlaps with a translation of its flip, hence it lies between $0$ and $1$. We then define the asymmetry value $\alpha$, which similarly lies in the same range. Under this definition, a circle will have an asymmetry value of $\alpha=0$, whereas a shape such as a triangle will have a higher asymmetry value. We illustrate \rev{our method of measuring pupil asymmetry in} \fref{fig:asymmetry_levels}.

\subsection{Is more asymmetry better?}
\label{sec: more_better}
As previously discussed, nearly symmetric pupils suffer from the similarity of PSFs $\vy$ and $\vy_*$ formed by $\vx$ and $\vx_*$, respectively. Although our later empirical results aim to demonstrate this for more general conditions, we now analytically demonstrate that more asymmetry is better in a limiting case. For the purposes of analysis, let $\mP$ take on scalar values and be decomposed into symmetric and asymmetric components as follows
\begin{equation}
    \mP = \mP_s + \epsilon \mP_a,
    \label{eq: pupil_decomp}
\end{equation}
where $\mP_s$ is symmetric and $\mP_a$ is asymmetric. Note that the geometry of the asymmetric component is fixed but weighted by $\epsilon$.

The following property demonstrates that separability increases with additional asymmetry.

\begin{property}
    For a pupil decomposed according to \eqref{eq: pupil_decomp} with $0 < \epsilon \ll 1$ along with $\mF \mP_s \vone$ and $\mF \mP_s \vphi$ as real vectors, then
    \begin{equation}
        \norm{\vy - \vy_*}^2 \approx 16\epsilon^2 \norm{\Im{(\mF \mP_s \vphi) \odot (\mF \mP_a \vone) - (\mF \mP_s \vone) \odot (\mF \mP_a \vphi)}}^2,
        \label{eq: result_analysis}
    \end{equation}
    when $\vx \approx 1 + j \vphi$.
\end{property}
\begin{proof}
    See supplementary.
\end{proof}

The property states that the separation between the PSF and its potential conjugate flip ambiguity increases with additional asymmetry. The term within the $\ell_2$ \rev{norm is a constant for a given phase and pupil and further} implies when $\vphi = \vzero$ then $\norm{\vy - \vy_*}^2 = 0$. Interestingly, \rev{a pupil with no symmetry} would also produce $\norm{\vy - \vy_*}^2 = 0$. Given our definition of symmetry, only a pupil $\mP = \vzero$ would have a vanishing symmetric component. Otherwise, there can always be a translation for which there exists one overlap at the point where there is a non-zero entry of $\mP$.

\section{Empirical Trends in Pupil Design}

\begin{figure}
    \centering
    \includegraphics[width=0.98\linewidth]{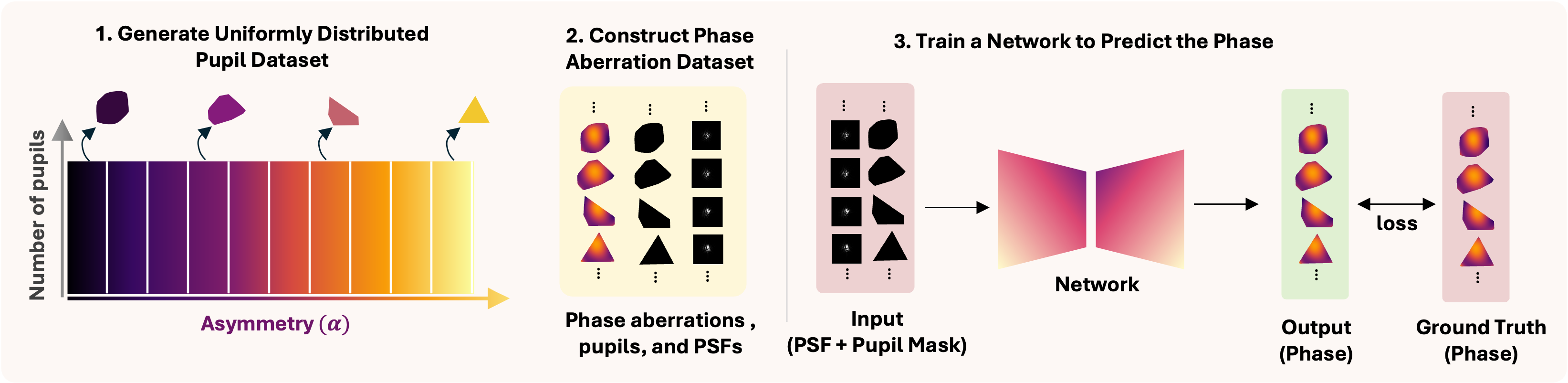}
    \caption{Overview of the pipeline of this paper. A dataset of pupils is generated to uniformly cover a range of asymmetry values. A large dataset of phase-pupil-PSF triplets is used to train a network in a supervised fashion.}
    \label{fig:pipeline}
\end{figure}

\subsection{Pupil and phase dataset Generation}
In this paper, we consider pupils that are convex hulls. A dataset of convex hulls with vertices ranging from 3 to 360 is generated to serve as our feasible set of pupils. We restrict our random pupils to have a minimum area within a reference circle to avoid small or near-delta pupils. To ensure that each asymmetry level is represented equally, we calculate the asymmetry of each pupil and bin them into 30 asymmetry ranges. Once a certain asymmetry level reaches the desired number of pupils, we reject any new random pupil with this level of asymmetry.

For a convex hull 2D shape, the maximum asymmetry is: ($\alpha = 0.\overline{3}$). This is derived from the symmetry measure known as the Kovner–Besicovitch measure and discussed by Grünbaum \cite{Grunbaum_symmetry}. Fáry later showed that the maximum asymmetric convex shape is a triangle \cite{Faray_triangle_asymmetry}. The maximum asymmetry we achieved ($\alpha \approx 0.35$) is slightly higher due to discretization.

\subsection{Different networks to mitigate network bias}
Although our dataset was designed to be uniformly distributed across asymmetry levels, and training performance was consistent across all asymmetries, indicating no bias toward any particular level, we also sought to verify that the observed recovery trend is not an artifact of a specific network architecture. To this end, we repeated the experiments using two architectures: a U-Net \cite{unet} and a fully connected MLP. Therefore, we can claim that not only is our dataset unbiased, but the observed recovery trend was also not a product of a specific choice of architecture. The results reported in the main paper use a U-Net, while MLP results can be found in the supplementary document.

\subsection{Light throughput and noise simulation}
Smaller pupils have smaller light throughput, and therefore in the presence of noise will have a lower signal-to-noise ratio (SNR). To account for this effect, we normalize all the PSFs generated by any pupil by the maximum of $\abs{\mF \mC \vone}^2$, then add noise. Mathematically, this is achieved as
\begin{equation}
    \vy = \frac{\abs{\mF \mP \vx}^2}{\max \abs{\mF \mC \vone}^2} + \vn,
\end{equation}
where $\vn \sim \calN(\vzero, \sigma^2 \mI)$  has a fixed variance for all the pupils. Smaller pupils will yield lower energy in the PSF and therefore, when normalized and noise is added, they would have much lower SNR than bigger pupils. This helps to keep the range of the PSFs reasonable for our network, but is a consistent normalization to reflect optical throughput.

\begin{figure}[H]
    \centering
    \includegraphics[width=0.8\linewidth]{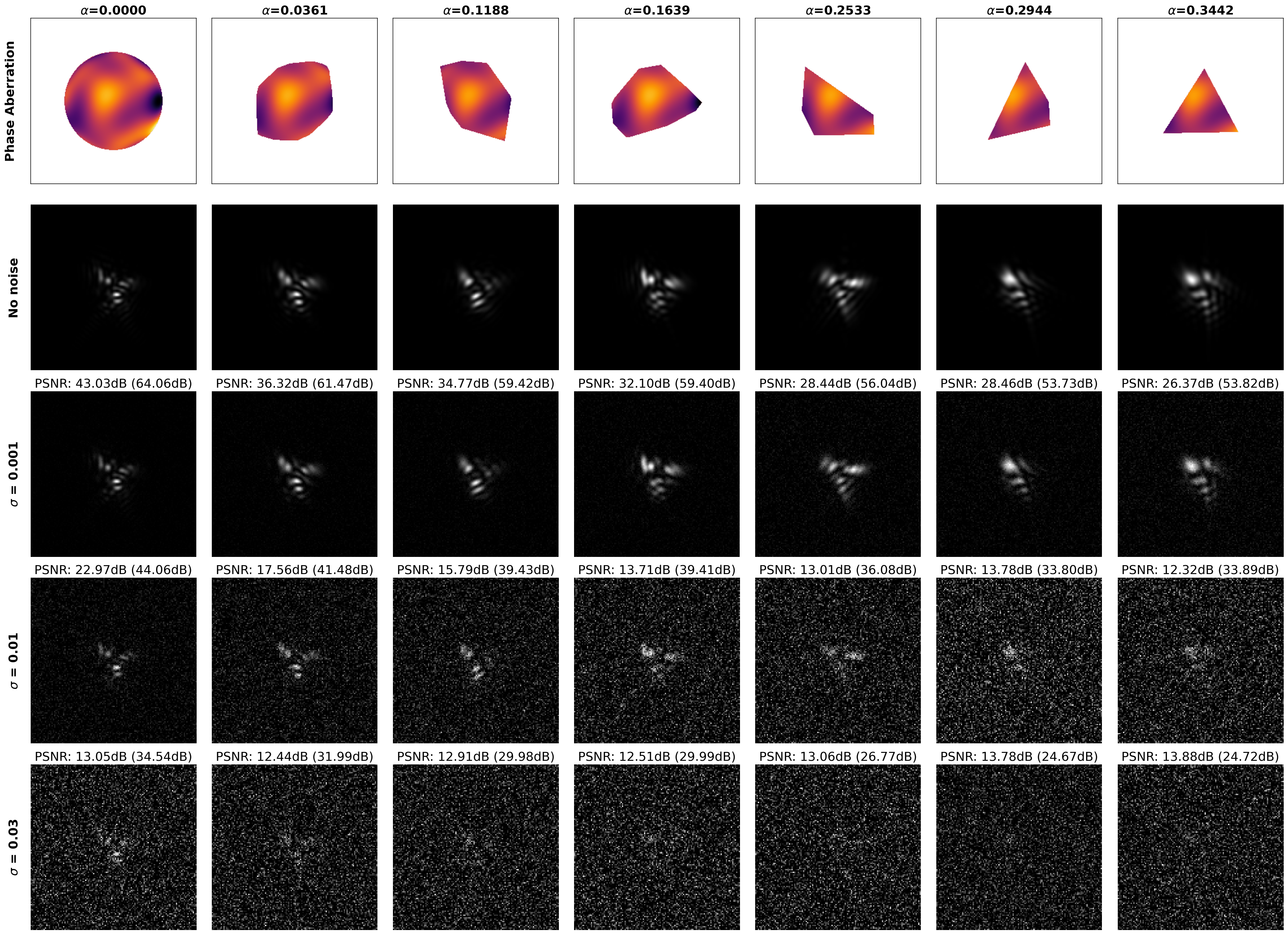}
    \caption{Qualitative light throughput and noise results. With our simulation, more asymmetric pupils yield lower PSNR even when the noise variance ($\sigma^{2}$) is fixed, reflecting a physics-grounded light throughput simulation. The PSNR of each PSF is noted, with the average PSNR for that specific pupil and noise sigma in parentheses.}
    \label{fig:PSF_grid}
\end{figure}

The noise variance $\sigma$ is chosen so that it yields a certain maximum PSNR with the diffraction-limited circular PSF $\vy_{c}$. While those PSNR values are high for the circular pupil PSF, they highly penalize the smaller non-circular pupils. We visualize the average PSNR across asymmetry levels in \fref{fig:PSF_grid}, which intuitively reduces \rev{as asymmetry increases and pupils shrink, resulting in a lower SNR}.

\subsection{Trends due to design criteria}

\subsubsection{Recovery versus asymmetry.} 
\begin{figure}[H]
\centering
    \begin{tabular}{cc}
    \centering
        \includegraphics[width=0.40\linewidth]{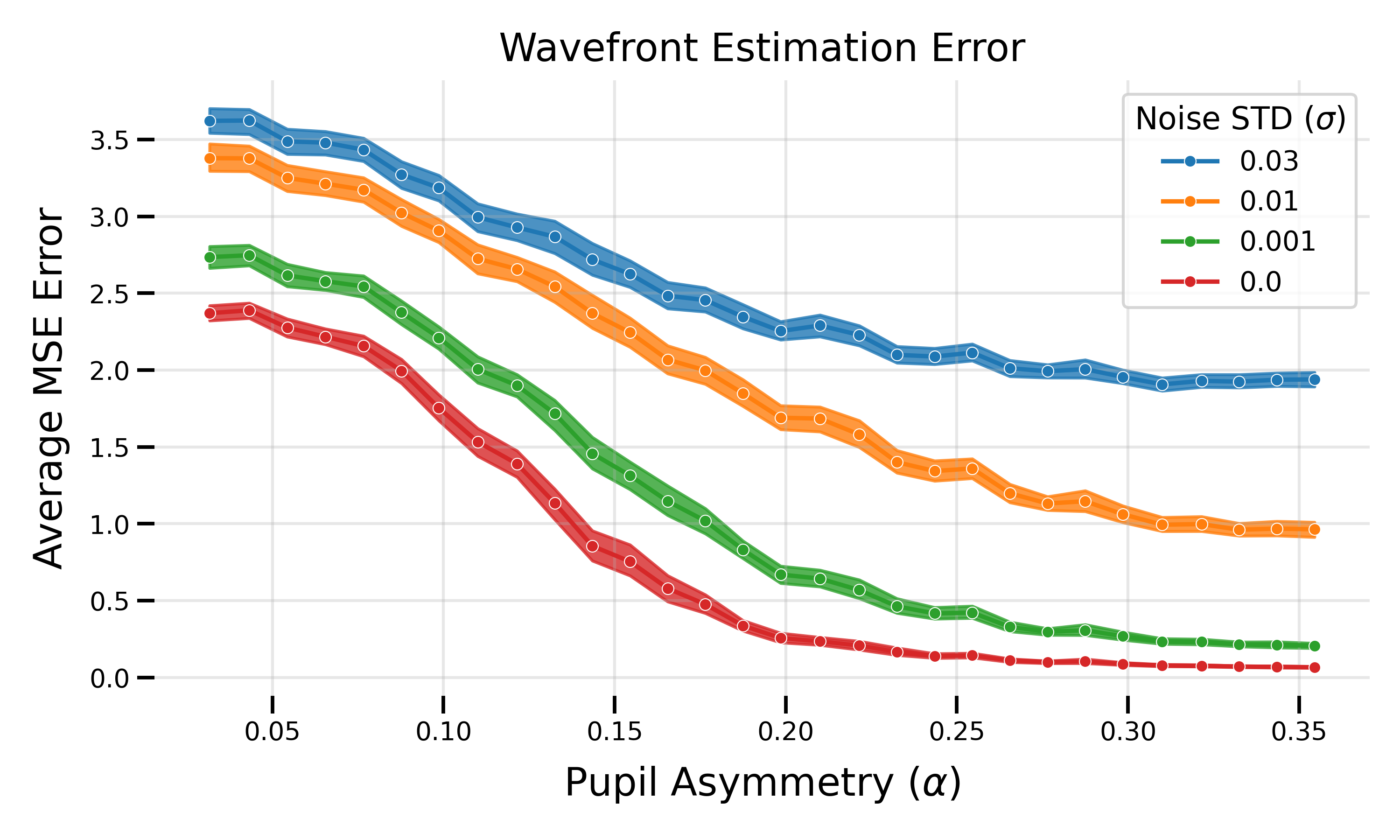} & \includegraphics[width=0.40\linewidth]{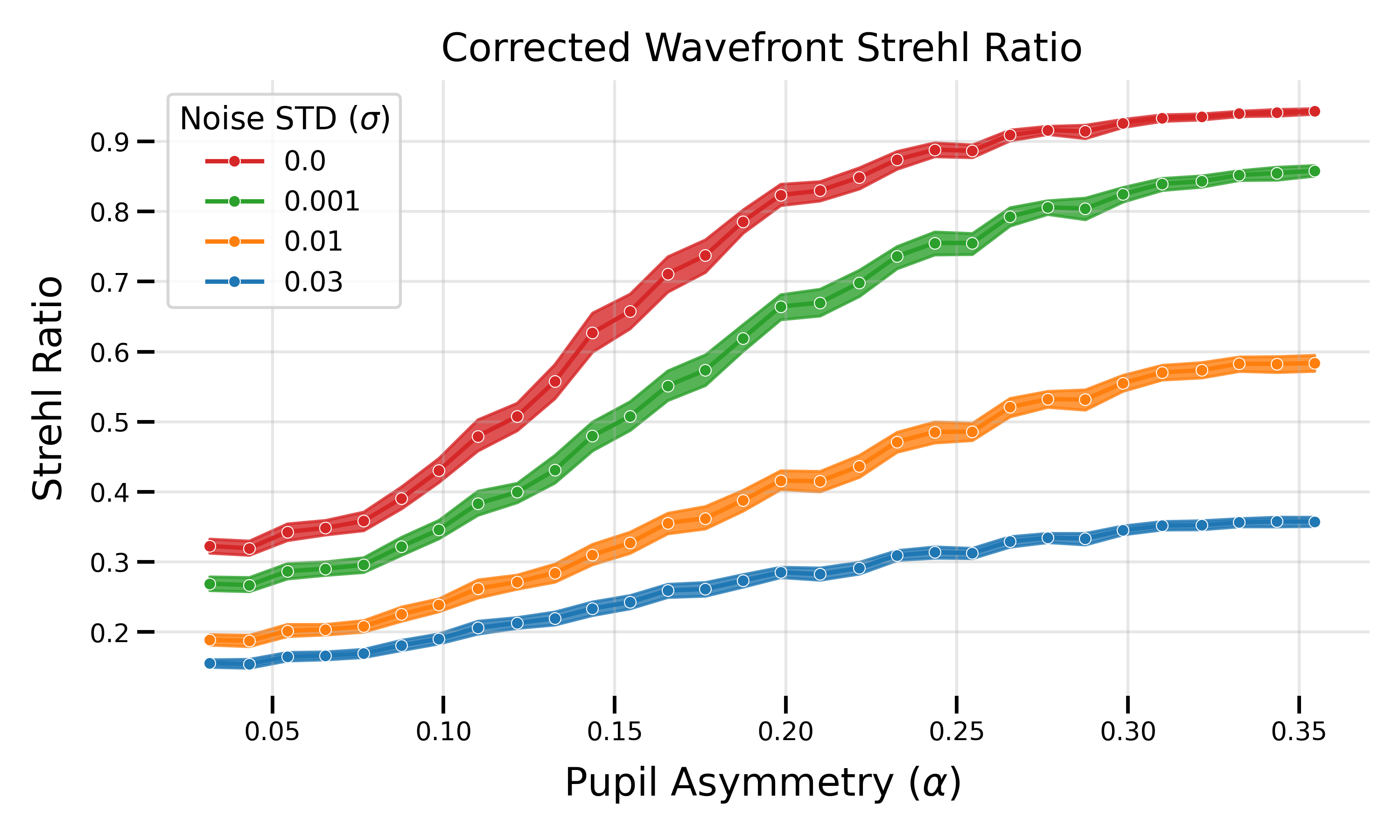} \\
        (a) Wavefront estimation loss & (b) Strehl ratio
    \end{tabular}   
    
    \caption{(a) Wavefront estimation MSE loss as a function of pupil asymmetry (lower is better), and (b) Corrected wavefront Strehl ratio as a function of pupil asymmetry (higher is better). Results reported in this figure use testing data.}
    \label{fig:general_trend}
\end{figure}

We conducted several experiments with increasing level or noise in the PSFs. Our experiments demonstrate that the more asymmetric pupils yield lower wavefront estimation error and higher Strehl ratio on testing data as seen in \fref{fig:general_trend}, supporting that the asymmetric metric directly relates to downstream recovery and serves as a suitable proxy for pupil design.

\begin{figure}[t!]
    \centering
    \includegraphics[width=0.95\linewidth]{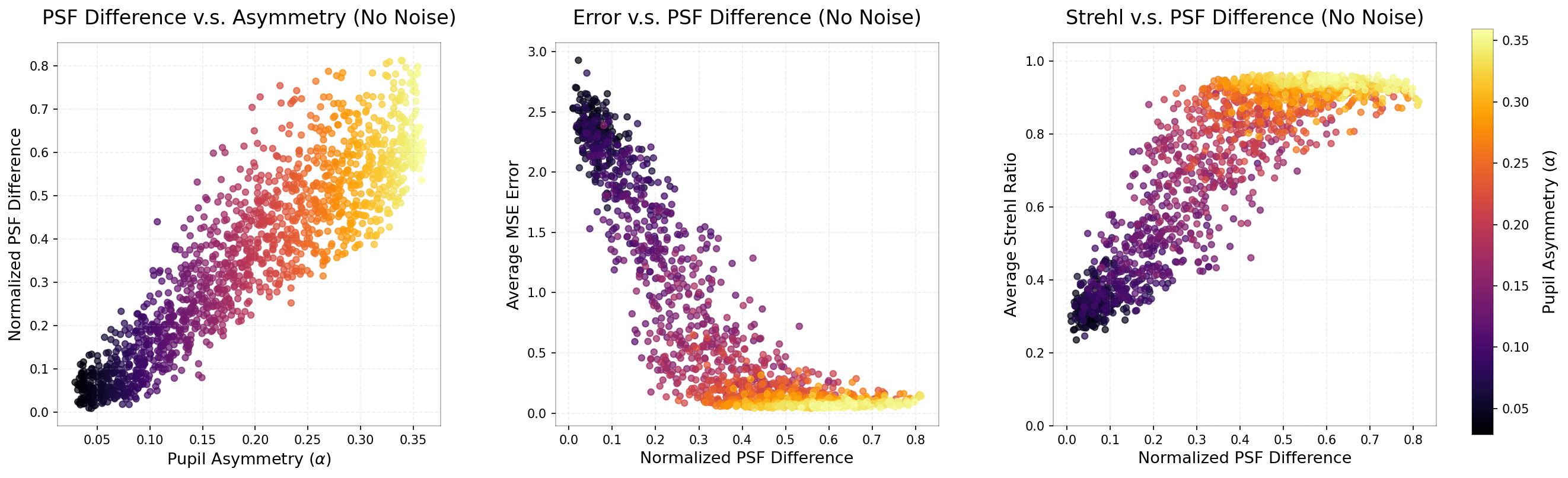}
    \caption{The normalized difference between the PSF and the PSF of the conjugate flip. More asymmetric pupils encode higher PSF difference, and that correlates with more distinguishability, and therefore lower wavefront estimation error, and higher Strehl ratios.}
    \label{fig:psf_difference}
\end{figure}

\subsubsection{Ambiguous PSF separation and asymmetry.} As discussed in Section \ref{sec: more_better}, the asymmetric part of the pupil contributes to the distinguishability between the PSF of the pupil field and the PSF of its conjugate flip, which translates to fewer ambiguous solutions the network has to choose between. We visualize the normalized difference between a PSF and the PSF of the conjugate flip in \fref{fig:psf_difference}. We observe a strong correlation between asymmetry and ambiguous PSF difference, which translates to lower wavefront estimation error and improved Strehl ratios.
\begin{figure}[H]
    \includegraphics[width=0.95\linewidth]{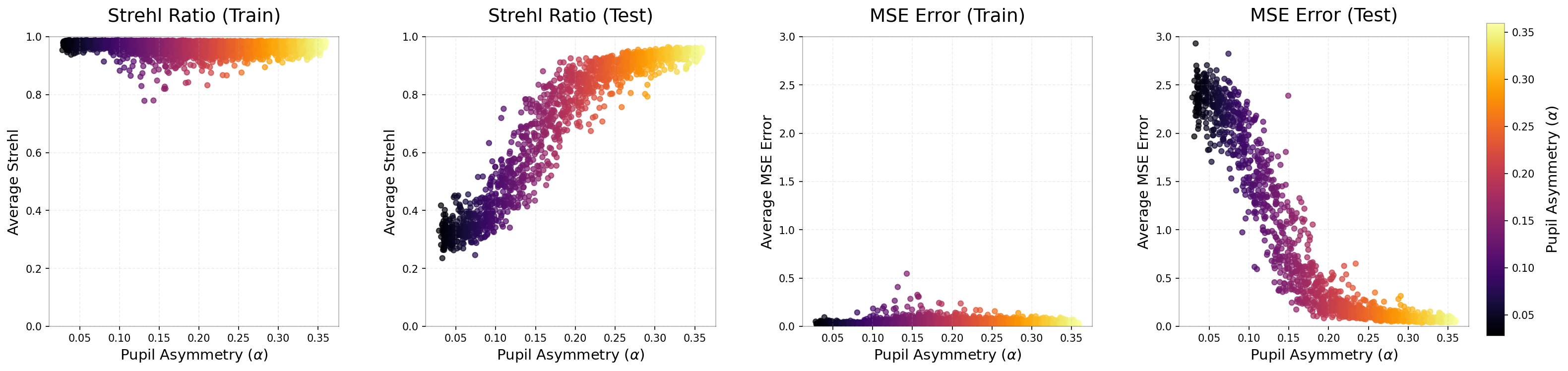}
    \caption{Mean Strehl ratio and MSE for noiseless training and testing sets. Each point represents a pupil performance averaged over 5000 (train) and 1500 (test) phase aberrations. A total of 30 million datapoints (6000 pupils $\times$ 5000 phases), and 1.5 million (1000 pupils $\times$ 1500 phases) were calculated for training and testing, respectively. Consistent training results across asymmetry levels confirm that the network is unbiased, while testing performance highlights the superior recoverability and generalizability of asymmetrical pupils.}
    \label{fig:training_and_testing_performance}
\end{figure}

\subsubsection{Network bias and generalization.} To further verify that the observed trend is not due to network bias toward one asymmetry level, we compare the network performance over all the entire training dataset with performance on the testing dataset. \fref{fig:training_and_testing_performance} shows a uniform flat loss and Strehl ratio across all asymmetries on the training data, which suggests that the training does not favor one type of asymmetry over another. However, generalization is negatively impacted for symmetric pupils by overfitting, related to our discussion in Section \ref{sec: overfitting_reason}.

\subsubsection{MSE versus Strehl ratio.} 
\begin{figure}[htbp]
    \centering
    \begin{subfigure}[b]{0.24\textwidth}
        \centering
        \includegraphics[width=\linewidth]{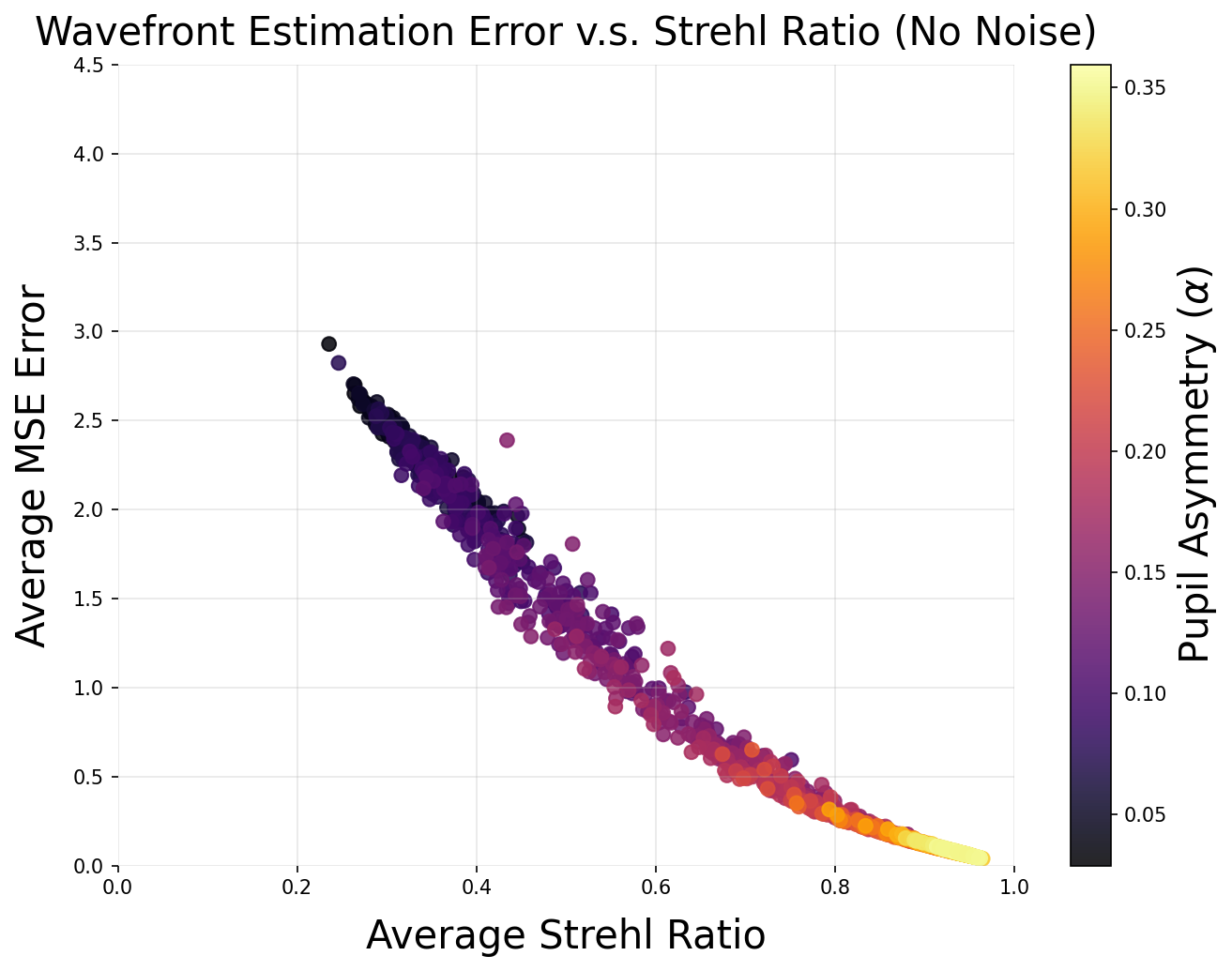}
        \caption{No noise}
        \label{fig:no_noise}
    \end{subfigure}
    \hfill
    \begin{subfigure}[b]{0.24\textwidth}
        \centering
        \includegraphics[width=\linewidth]{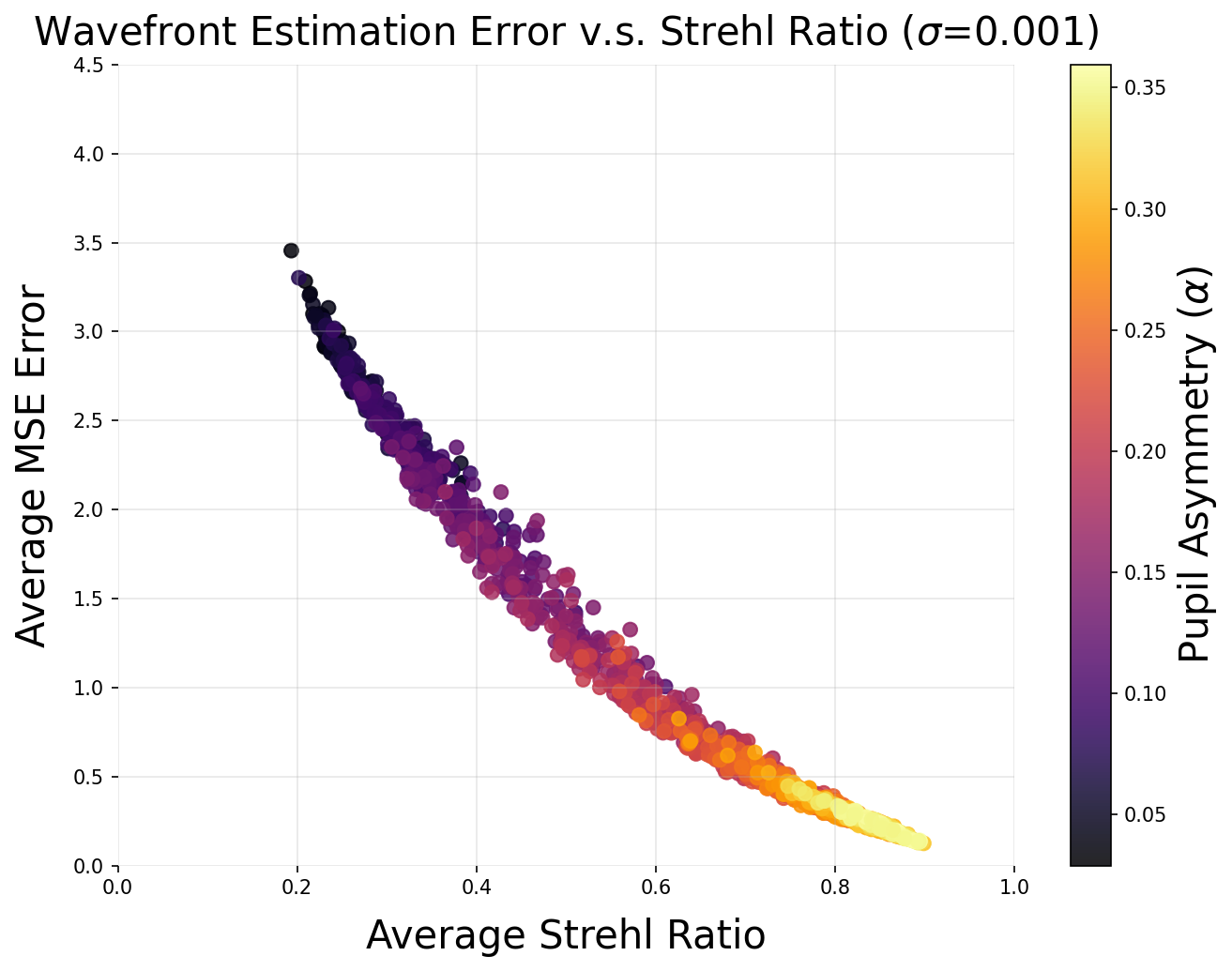}
        \caption{$\sigma = 0.001$}
        \label{fig:noise_001}
    \end{subfigure}
    \hfill
    \begin{subfigure}[b]{0.24\textwidth}
        \centering
        \includegraphics[width=\linewidth]{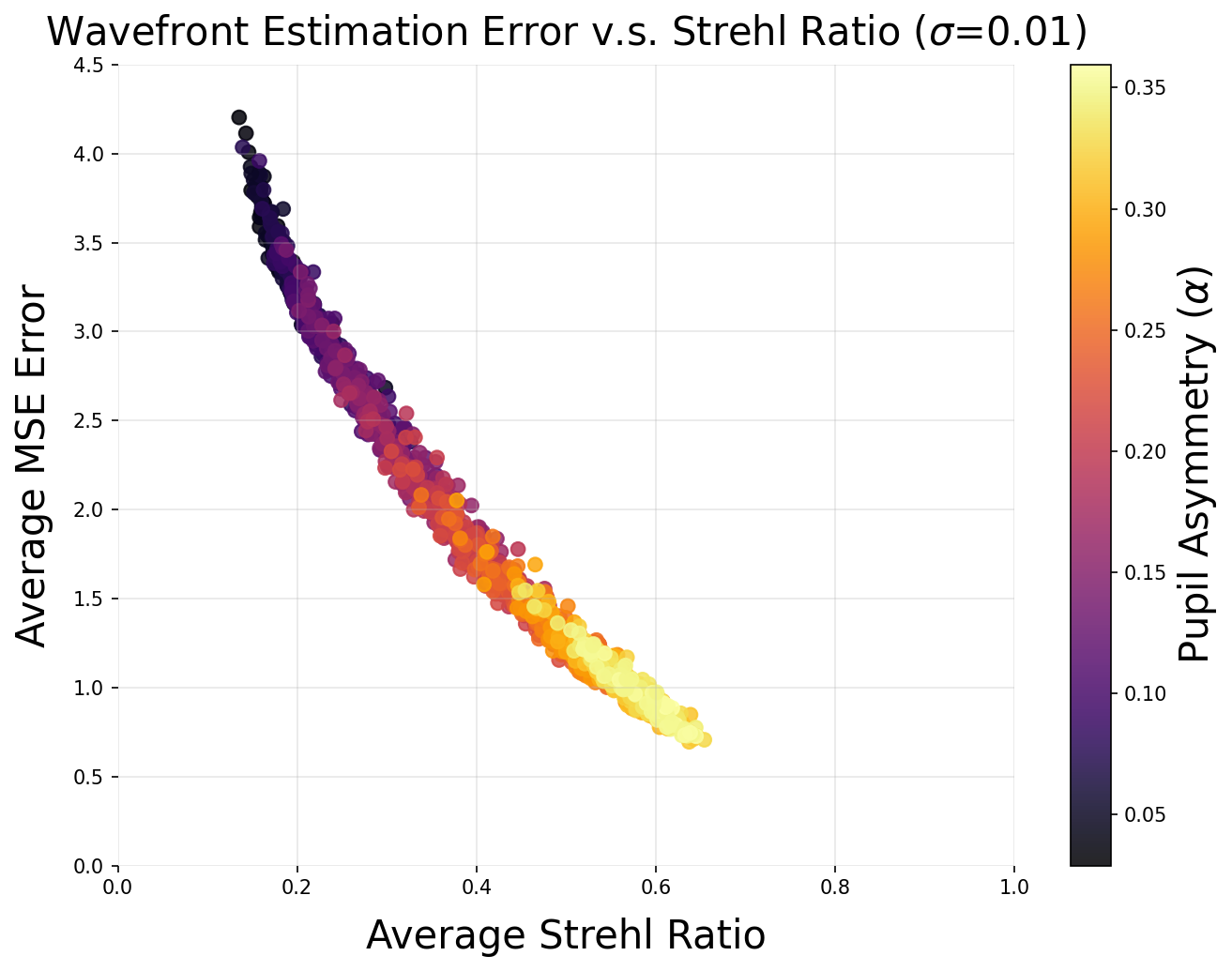}
        \caption{$\sigma = 0.01$}
        \label{fig:noise_01}
    \end{subfigure}
    \hfill
    \begin{subfigure}[b]{0.24\textwidth}
        \centering
        \includegraphics[width=\linewidth]{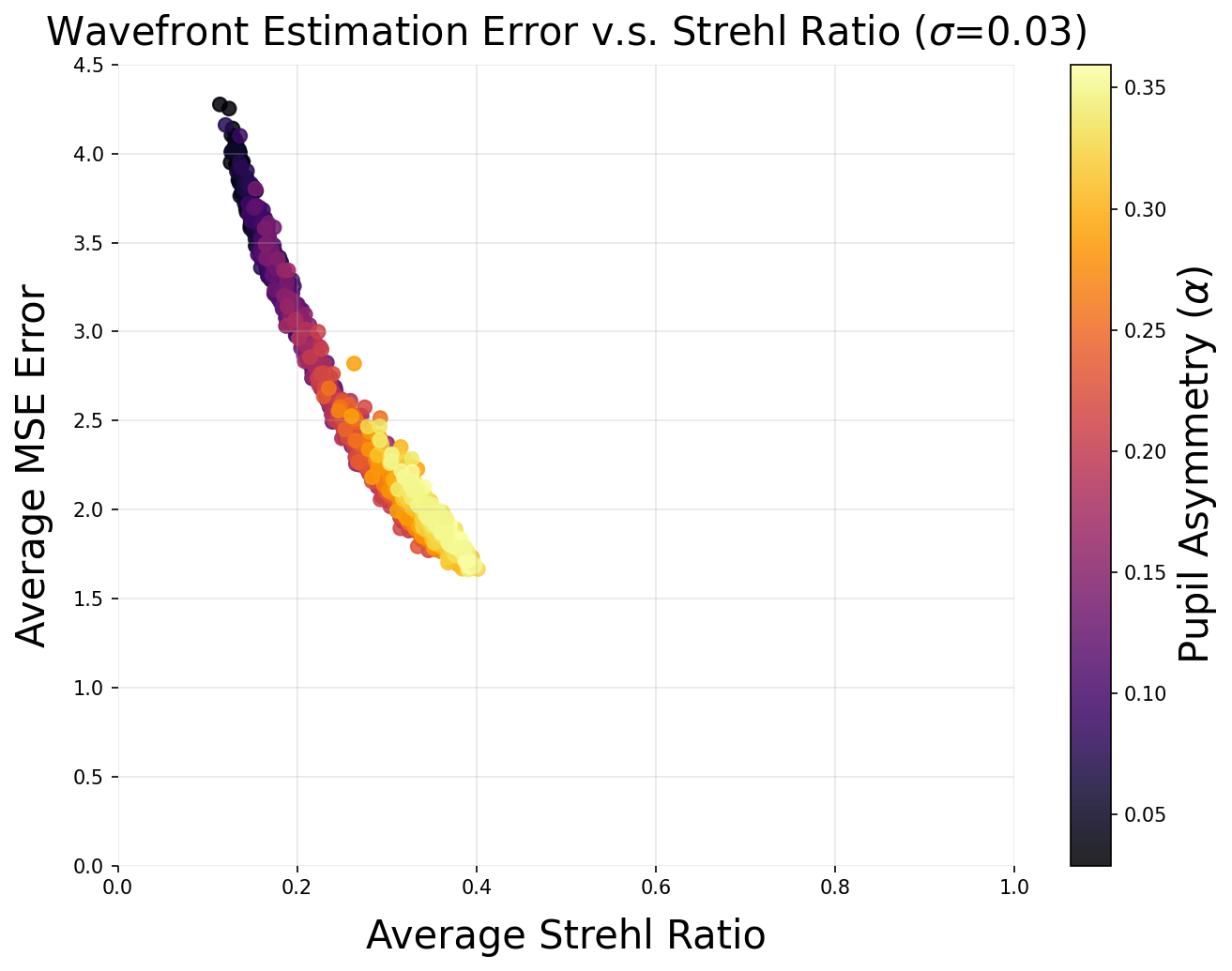}
        \caption{$\sigma = 0.03$}
        \label{fig:noise_03}
    \end{subfigure}

    \caption{Visualization of pupil asymmetry effect on wavefront estimation loss and the Strehl ratio of the corrected wavefront. More asymmetry yields lower error and a higher Strehl ratio, implying minimizing the MSE loss is a meaningful objective function to maximize Strehl ratio. The loss increases and Strehl decreases with more noise, but the asymmetry trend is consistent. All plots are on the same scale for easier interpretation. The colors represent increasing asymmetry.}
    \label{fig:strehl_vs_loss_vs_asymmetry}
\end{figure}

We further verify that the MSE objective function is suitable to optimize the Strehl ratio by plotting the MSE error versus the Strehl ratio per pupil in \fref{fig:strehl_vs_loss_vs_asymmetry}. The strong negative correlation implies that reducing MSE error is comparable to increasing the Strehl ratio, which holds across different noise levels.

\subsubsection{Aberration strength and performance.} 
\begin{figure}[h]
    \centering
    \begin{subfigure}{0.48\linewidth}
        \centering
        \includegraphics[width=\linewidth]{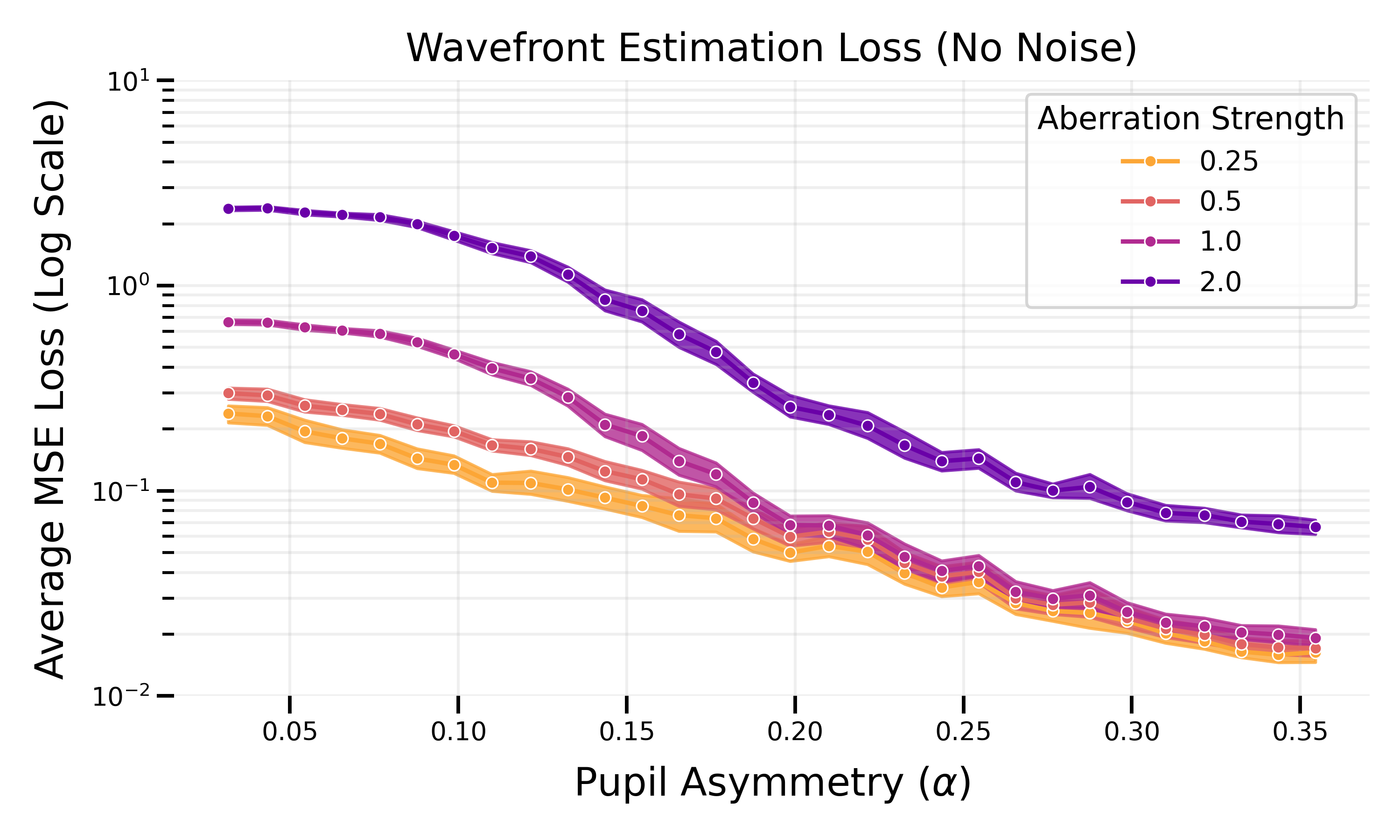}
        \caption{Wavefront Recovery MSE (No noise)}
    \end{subfigure}
    \hfill
    \begin{subfigure}{0.48\linewidth}
        \centering
        \includegraphics[width=\linewidth]{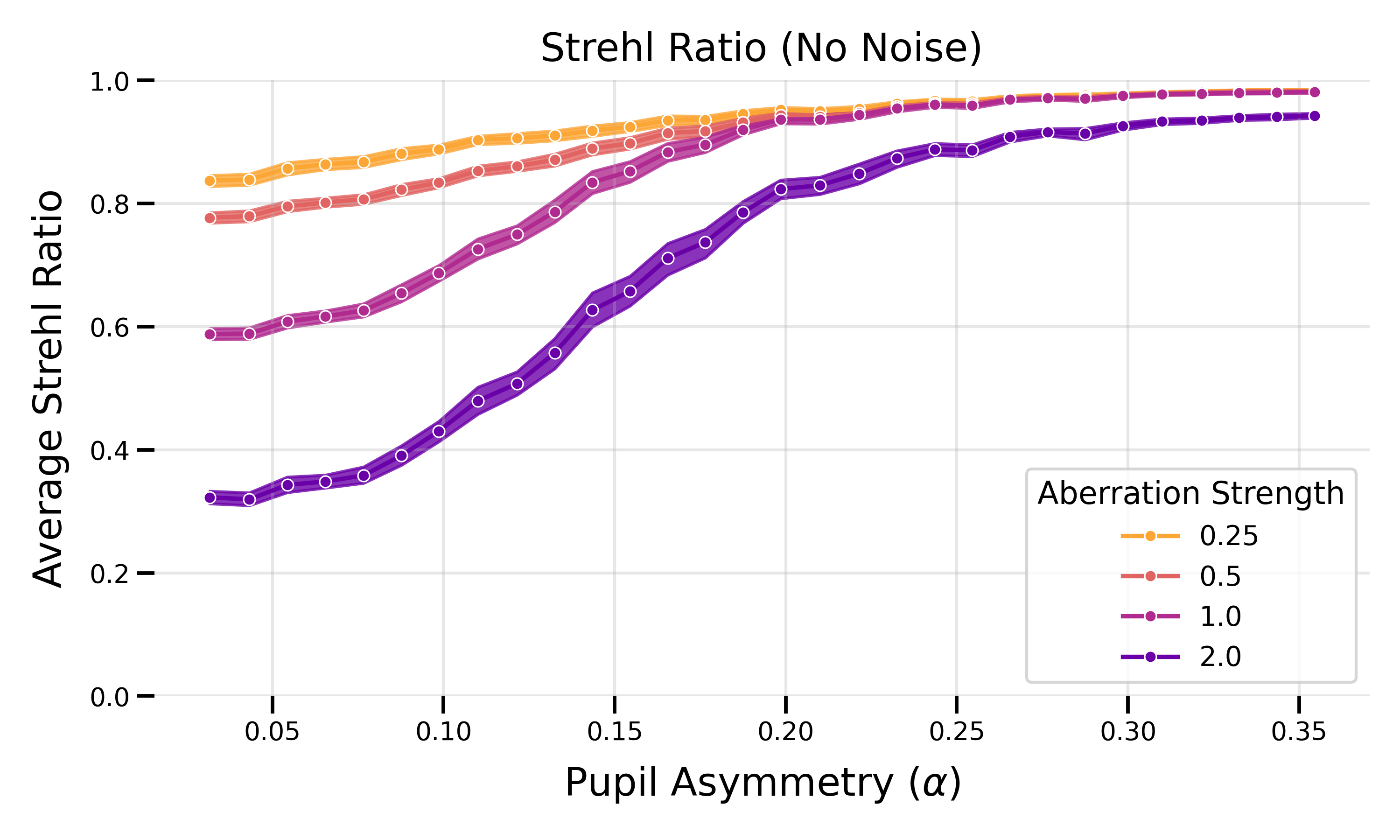}
        \caption{Strehl Ratio (No noise)}
    \end{subfigure}

    \caption{Comparison of wavefront estimation performance in different aberration scenarios (0.25 to 2.0) from the noise-free case. As the aberration scale increases, symmetric pupils exhibit higher error and degradation in the Strehl ratio. The MSE is shown in log scale for visibility between different aberration strengths.}
    \label{fig:multi_experiment_scale_comparison}
\end{figure}

The strength of aberration affects the wavefront recovery performance. In these experiments, we test pupil asymmetry against different levels of aberrations. Higher aberration strengths are achieved by increasing the Zernike coefficient range of values. Our experiments suggest that stronger aberrations result in degraded performance for the symmetric pupils. Therefore, the stronger the aberrations, the bigger the performance gap between the low asymmetry and high asymmetry pupils as shown in \fref{fig:multi_experiment_scale_comparison}.

\subsection{Single system performance}
The corrected wavefront Strehl ratios presented in the previous section depict the correction abilities of different pupils. These results were calculated relative to the pupil itself, i.e., using $\rho_{\mP}$. We now quantify the trade-off introduced by reducing the optical throughput and reducing the optical resolution by the reductive form of asymmetry used in this paper. This represents the most pessimistic view of the trade-off between wavefront correction and the entire system's optical performance.

\begin{figure}[htbp]
    \centering
    \includegraphics[width=0.5\linewidth]{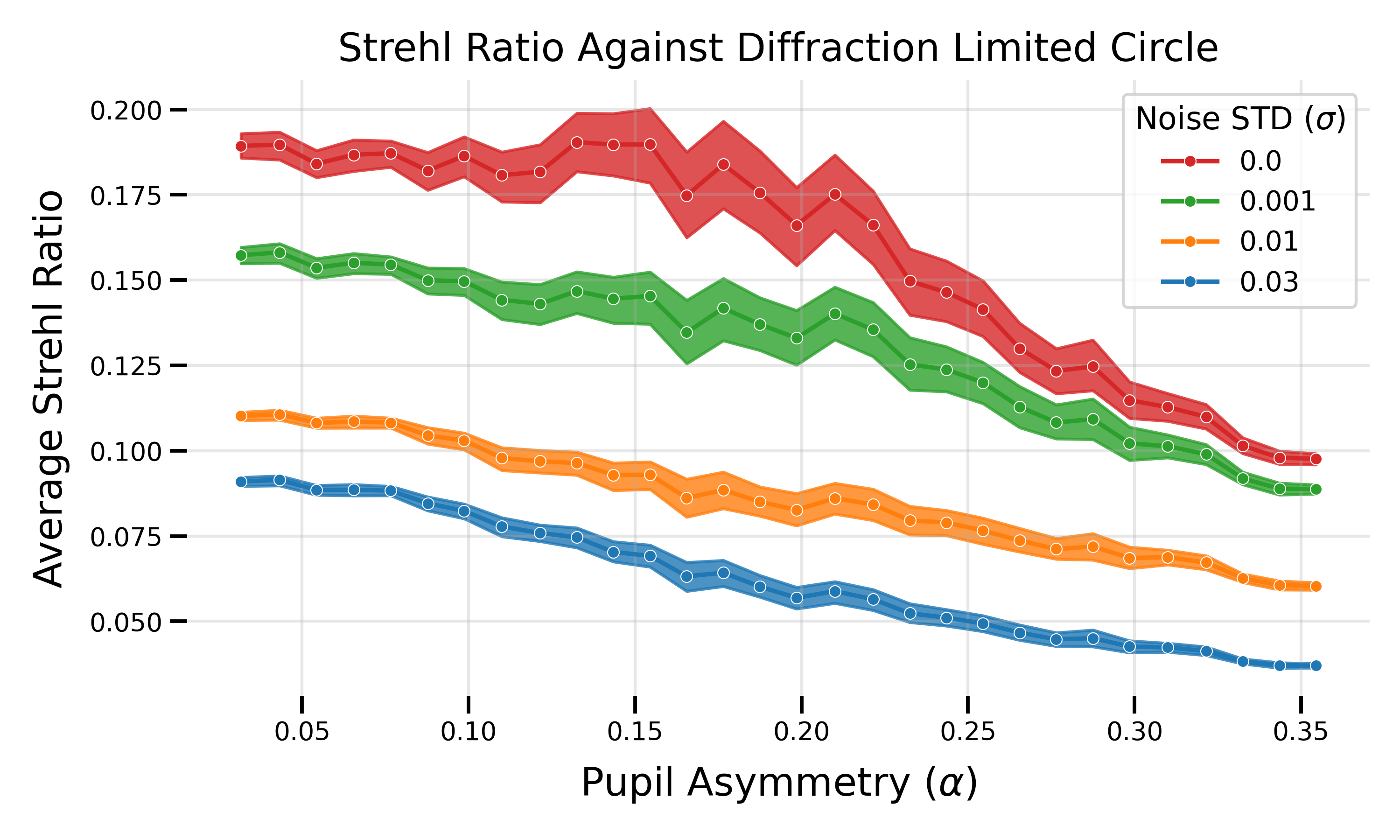}
    \caption{The Strehl ratio against the reference circular pupil diffraction limit in the presence of different measurement noise.}
    \label{fig:strehl_against_circle}
\end{figure}

We compute the Strehl ratio against the diffraction limit of the reference circular pupil according to \eqref{eq: strehl}. This ratio can be thought of as encompassing both the wavefront correction error and the inherent diffraction limit penalty by reducing pupil size to introduce asymmetry. We present the results in \fref{fig:strehl_against_circle}. These results indicate that for a single system performing both wavefront estimation and correction, performance gains may be limited. However, we emphasize that this is the most pessimistic analysis possible; if one can \textit{add} asymmetry to the pupil rather than achieve asymmetry by reduction, the performance trade-off will lie between the results of \fref{fig:strehl_against_circle} and those presented previously.

\begin{figure}[htbp]
    \centering
    \includegraphics[width=0.95\linewidth]{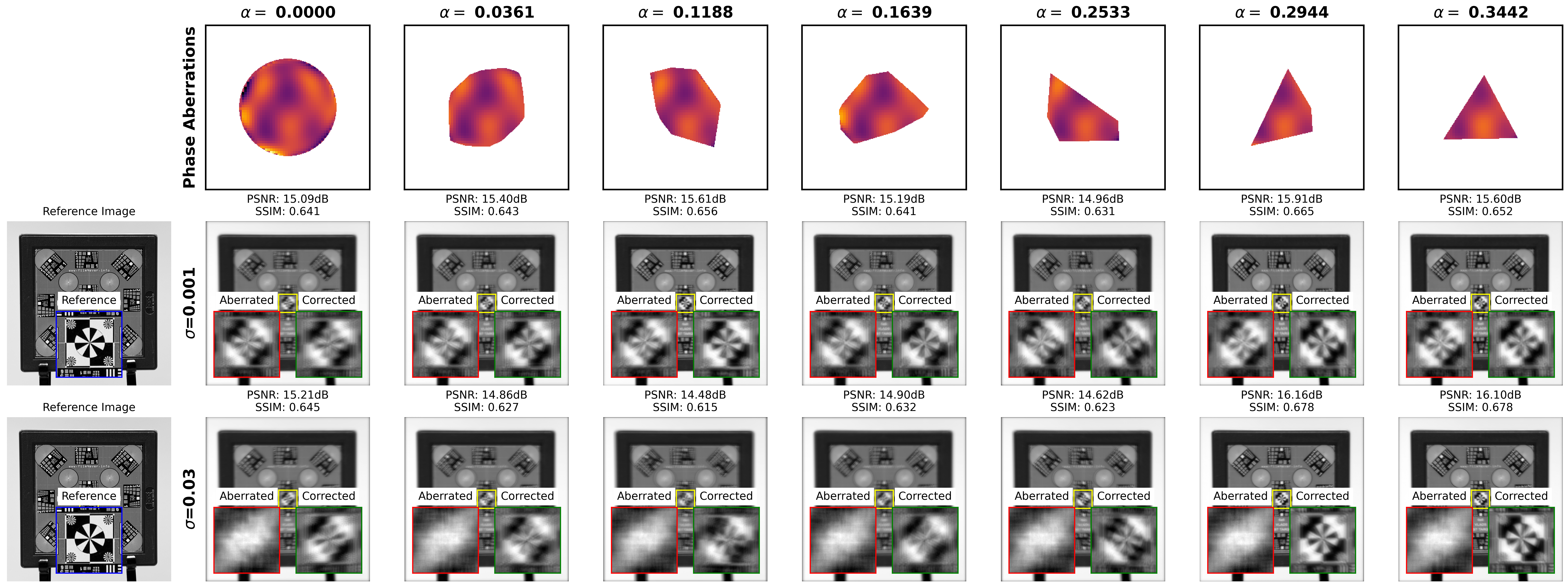}
    \caption{Wavefront correction simulated results with low and high noise. The target is blurred with the aberration PSF produced by the phase in the first row. The corrected image is shown for different levels of noise, with the aberrated and corrected images in the left and right insets, respectively. The image in the background is the corrected version.}
    \label{fig:correction_results}
\end{figure}

To visualize the effect of aberration correction on natural scenes, we show an example in \fref{fig:correction_results} where an image is synthetically aberrated with a PSF, and synthetically corrected with the phase predicted by the network. Pupils with more asymmetry tend to outperform those with less, especially in the presence of noise.

\section{Real Data Collection and Experiments}
To validate our results further, we built a prototype optical system to test the performance of different pupil asymmetries on wavefront recoverability. This section depicts the system, discusses the technical setup details, and reports the results. 

\subsection{Optical setup}
\begin{figure}[htbp]
    \begin{subfigure}{0.3\linewidth}
        \centering
        \includegraphics[width=\linewidth]{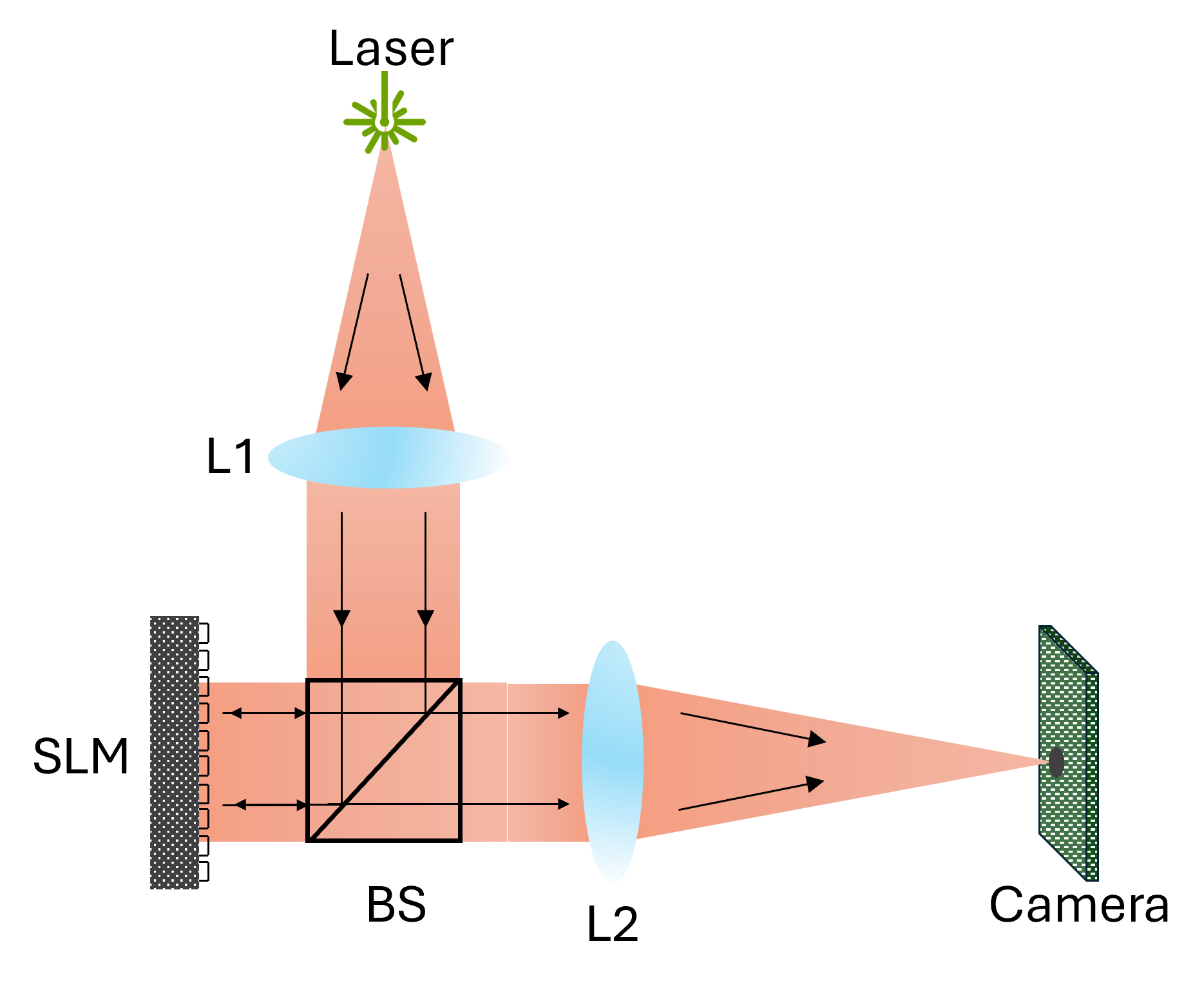}
        \caption{Optical setup diagram.}
    \end{subfigure}
    \begin{subfigure}{0.3\linewidth}
        \centering
        \includegraphics[width=\linewidth]{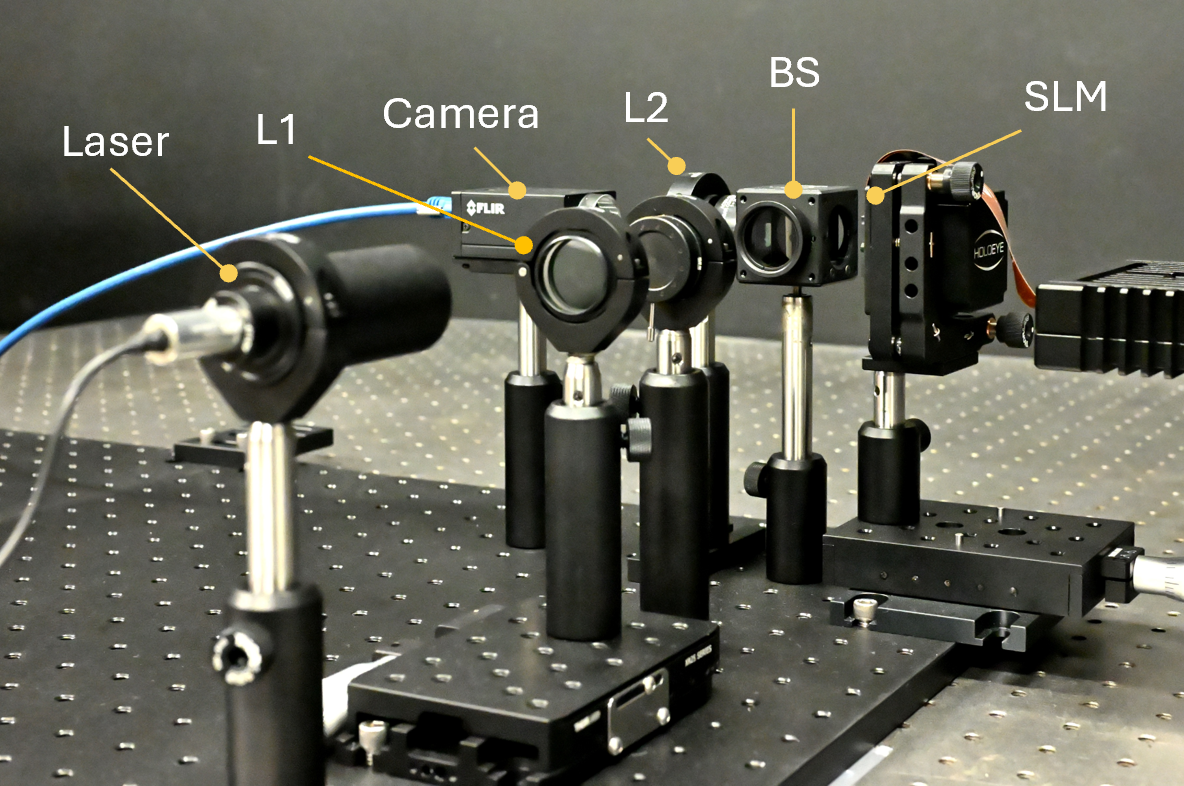}
        \caption{Corresponding setup.}
    \end{subfigure}
    \begin{subfigure}{0.3\linewidth}
        \centering
        \includegraphics[width=\linewidth]{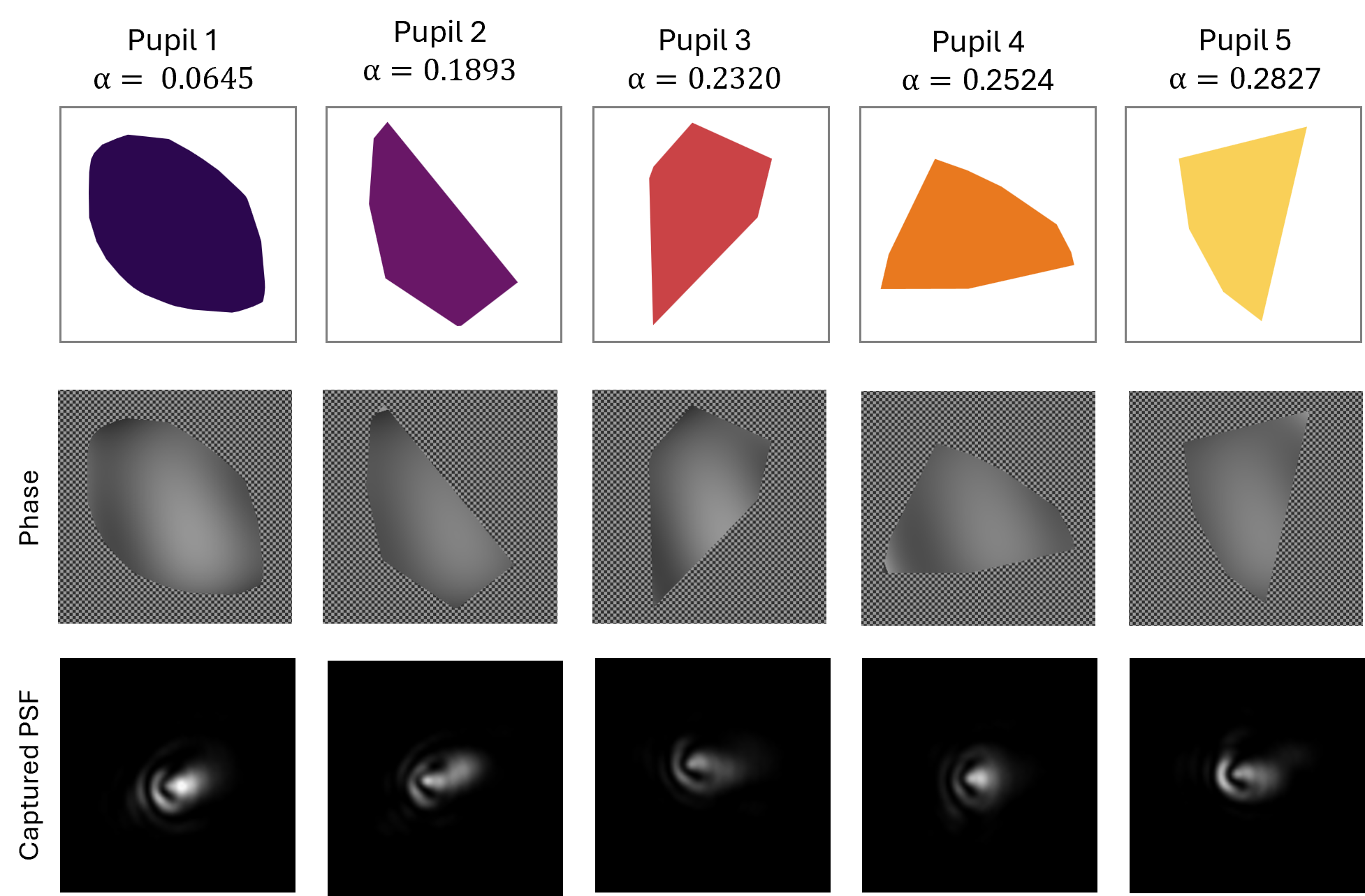}
        \caption{Real measurements.}
    \end{subfigure}
    \caption{The collimated light source illuminates the SLM. The reflected light passes through the focusing lens. The PSF is then captured by a camera. L1 is the collimating lens. L2 is the focusing lens. BS is a 50:50 beamsplitter. Examples from the real data acquisition (c) with phase pattern (second row) and the captured PSFs (third row).}
    \label{fig:optical system}
\end{figure}
Our optical setup is illustrated in \fref{fig:optical system}. We use a 520nm laser as the light source. The light beam is first collimated by lens L1 ($f=125mm$) and passes through a polarizer to ensure the appropriate polarization state required by the HOLOEYE GAEA-2.1 phase-only spatial light modulator (SLM). Phase aberrations are loaded onto the SLM. To realize arbitrary asymmetric pupils, a checkerboard pattern is displayed as the background of the SLM, enforcing zero amplitude outside the pupil region \cite{mendoza2014encoding}. The focusing lens L2 ($f=75mm$) performs the spatial Fourier transform of the complex wavefront reflected from the SLM. The camera (Flir Grasshopper3) is positioned in the Fourier plane of L2 to capture the point spread functions.

For our real experiments, we use five different pupils ranging in asymmetry. We randomly generated 500 phase patterns and used them for each pupil. The phases are wrapped to $[-\pi,\pi]$. In total, 2500 PSFs were captured for the real experiments. Representative examples for each pupil are shown in \fref{fig:optical system}(c).

\subsection{Real experiment results}
We evaluate our real data experiment by training a U-Net model to predict the phase aberrations from the captured PSFs. The input to the model is the PSF and the binary pupil mask, consistent with our previous experiments, and is trained to minimize the MSE. The results from our real experiment show a similar trend in performance versus asymmetry as shown in Table \ref{tab:real_experiment_results}.

\begin{table}[htbp]
    \centering
    \caption{Real data average wavefront estimation error on a testing subset. The more asymmetric the pupil, the lower the error. Highest asymmetry and lowest error are highlighted in boldface. Corresponding pupils are shown in \fref{fig:optical system}.}
    \label{tab:pupil_performance}
        \begin{tabular}{lccc}
        \hline
        \textbf{Pupil} & \textbf{Asymmetry ($\alpha$)} & \textbf{RMSE (rad)} & \textbf{MSE (rad$^2$)} \\ \hline
        Pupil 1 & 0.0654 & 0.379285 & 0.143857\\
        Pupil 2 & 0.1893 & 0.353183 & 0.124739\\
        Pupil 3 & 0.2320 & 0.323678 & 0.104768\\
        Pupil 4 & 0.2524 & 0.315149 & 0.099319\\
        Pupil 5 &\textbf{ 0.2827 } & \textbf{0.305383 } & \textbf{0.093259} \\\hline
        \end{tabular}
    \label{tab:real_experiment_results}
\end{table}

\section{Conclusion}
Our work establishes a quantitative link between pupil design and wavefront recoverability by introducing a formal asymmetry metric. Through extensive simulated and real experiments, we demonstrated the trade-offs between pupil asymmetry, wavefront recovery, and correction in different noise and aberration strength regimes.  This research provides guidance for designing pupils that enable unambiguous wavefront estimation from a single measurement, a task essential to many imaging applications, from adaptive optics to computational microscopy and beyond, enabling real-time phase estimation and correction.

\bibliographystyle{splncs04}
\bibliography{main}

\clearpage

\setcounter{section}{0}
\setcounter{figure}{0}
\setcounter{table}{0}
\setcounter{equation}{0}

\renewcommand{\thesection}{S\arabic{section}}
\renewcommand{\thefigure}{S\arabic{figure}}
\renewcommand{\thetable}{S\arabic{table}}
\renewcommand{\theequation}{S\arabic{equation}}

\title{Pupil Design for Computational Wavefront Estimation \texorpdfstring{\\}{ } (Supplementary Material)}

\author{Ali Almuallem\inst{1}\orcidlink{0000-0002-5235-077X} \and
Nicholas Chimitt\inst{1}\orcidlink{0000-0001-9528-0102} \and
Bole Ma\inst{1}\orcidlink{0009-0007-0417-3979} \and
Qi Guo\inst{3}\orcidlink{0000-0002-8329-7668} \and
Stanley H. Chan\inst{3}\orcidlink{0000-0001-5876-2073}}

\authorrunning{A. Almuallem et al.}

\institute{Elmore Family School of Electrical and Computer Engineering, Purdue University, USA}

\maketitle
\section{Proof of Property 1.}
\label{sec:epsilon}
We now provide a proof for Property 1 presented in the main paper. We recall that $\vy = \abs{\mF \mP \vx}^2$ and $\vy_* = \abs{\mF \mP \vx_*}^2$ and we further consider the pupil to take on scalar values and be decomposed as
\begin{equation}
    \mP = \mP_s + \epsilon \mP_a,
    \label{eq: pupil_decomp_appendex}
\end{equation}
where $\mP_s$ is symmetric and $\mP_a$ is asymmetric. Note that the geometry of the asymmetric component is fixed but weighted by $\epsilon$. The following property demonstrates that separability increases with additional asymmetry.
\begin{property}
    For a pupil decomposed according to \eqref{eq: pupil_decomp_appendex} with $0 < \epsilon \ll 1$ along with $\mF \mP_s \vone$ and $\mF \mP_s \vphi$ as real vectors, then
    \begin{equation}
        \norm{\vy - \vy_*}^2 \approx 16\epsilon^2 \norm{\Im{(\mF \mP_s \vphi) \odot (\mF \mP_a \vone) - (\mF \mP_s \vone) \odot (\mF \mP_a \vphi)}}^2,
        \label{eq: result_analysis_appenex}
    \end{equation}
    when $\vx \approx 1 + j \vphi$.
\end{property}
\begin{proof}
We begin by explicitly writing
\begin{align*}
    \mF \mP \vx &= \underbrace{\mF \mP_s \vone}_{\va} + \epsilon \underbrace{\mF \mP_a \vone}_{\vc} + j \underbrace{\mF \mP_s \vphi}_{\vb} + \epsilon j \underbrace{\mF \mP_a \vphi}_{\vd}.
\end{align*}
Furthermore, $\mF \mP \vx_* = \va + \epsilon \vc - j \vb - \epsilon j \vd$ since $\vx_* \approx 1 - j \vphi$ by small angle approximation. The vectors $\va$ and $\vb$ are real by assumption, while $\vc$ and $\vd$ are generally complex. Writing $\vc = \vc_r + j \vc_i$ and $\vd = \vd_r + j \vd_i$ where $\vc_r = \Re{\vc}$, $\vc_i = \Im{\vc}$ and similarly for $\vd$, then
\begin{align*}
    \mF \mP \vx &= (\va + \epsilon \vc_r - \epsilon \vd_i) + j(\vb + \epsilon \vc_i + \epsilon \vd_r), \\
    \mF \mP \vx_* &= (\va + \epsilon \vc_r + \epsilon \vd_i) + j(-\vb + \epsilon \vc_i - \epsilon \vd_r),
\end{align*}
where we have organized terms into real and imaginary components.
By $\abs{\mF \mP \vx}^2 = (\mF \mP \vx) \odot (\mF \mP \vx)^*$, then
\begin{align*}
    \vy &= (\va + \epsilon \vc_r - \epsilon \vd_i)^2 + (\vb + \epsilon \vc_i + \epsilon \vd_r)^2, \\
    \vy_* &= (\va + \epsilon \vc_r + \epsilon \vd_i)^2 + (-\vb + \epsilon \vc_i - \epsilon \vd_r)^2,
\end{align*}
where $(\cdot)^2$ corresponds to a Hadamard product $\odot$ as the elements are vectors. The difference between these two vectors can be shown to give
\begin{equation}
    \vy - \vy_* = -4(\va + \epsilon \vc_r) \odot (\epsilon \vd_i) + 4 (\vb + \epsilon \vd_r) \odot (\epsilon \vc_i).
\end{equation}
Dropping terms that depend on $\epsilon^2$, 
\begin{equation}
    \vy - \vy_* \approx 4\epsilon (\vb \odot \vc_i - \va \odot \vd_i)
\end{equation}
which can be further reduced to $4 \epsilon \Im{\vb \odot \vc - \va \odot \vd}$ by the fact that $\vb, \va$ are real. Replacing $\va, \vb, \vc, \vd$ by their definitions and taking the norm square completes the proof.
\end{proof}

\section{Additional MLP Performance}
\label{sec:mlp_results}
\begin{figure}[htbp]
\centering
    \begin{tabular}{cc}
    \centering
        \includegraphics[width=0.40\linewidth]{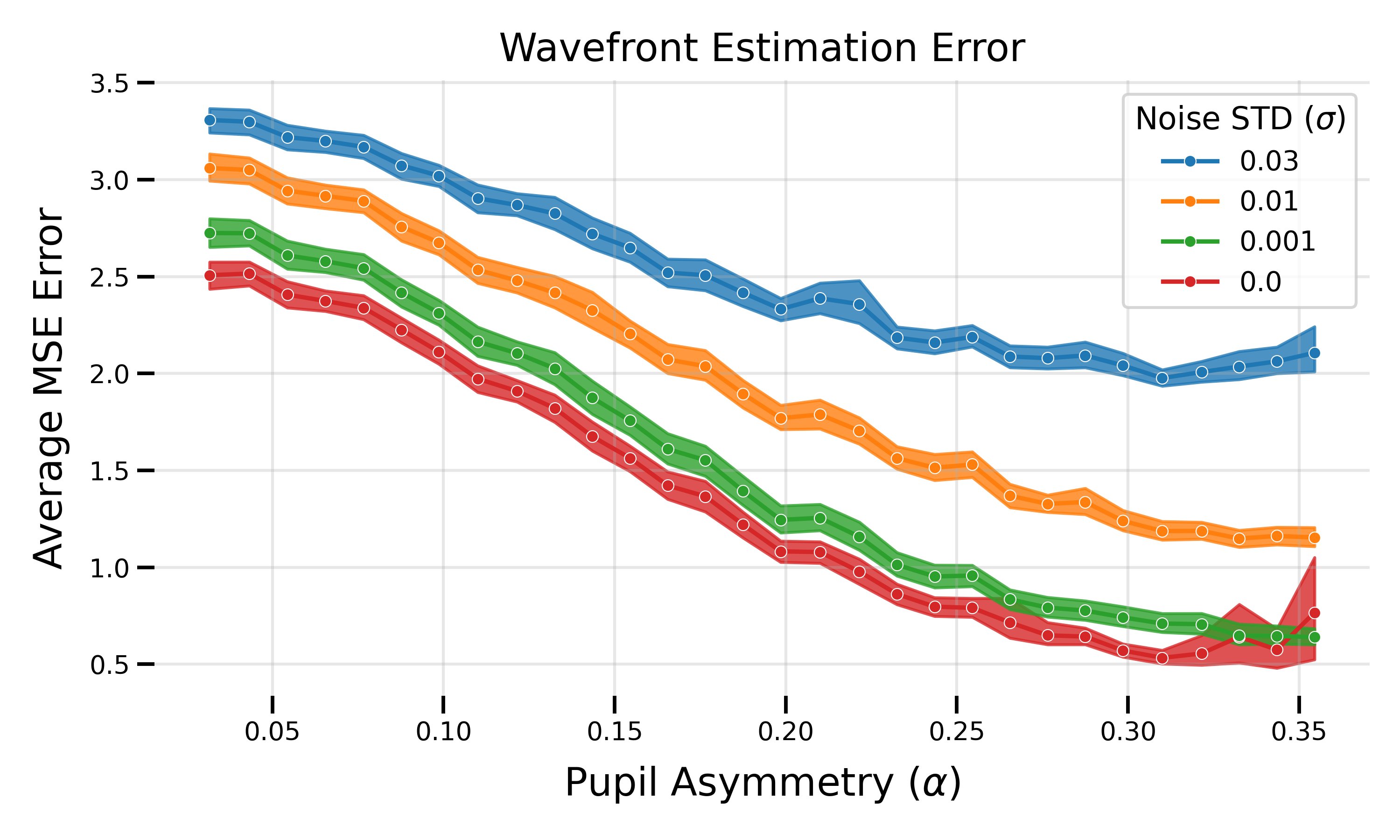} & \includegraphics[width=0.40\linewidth]{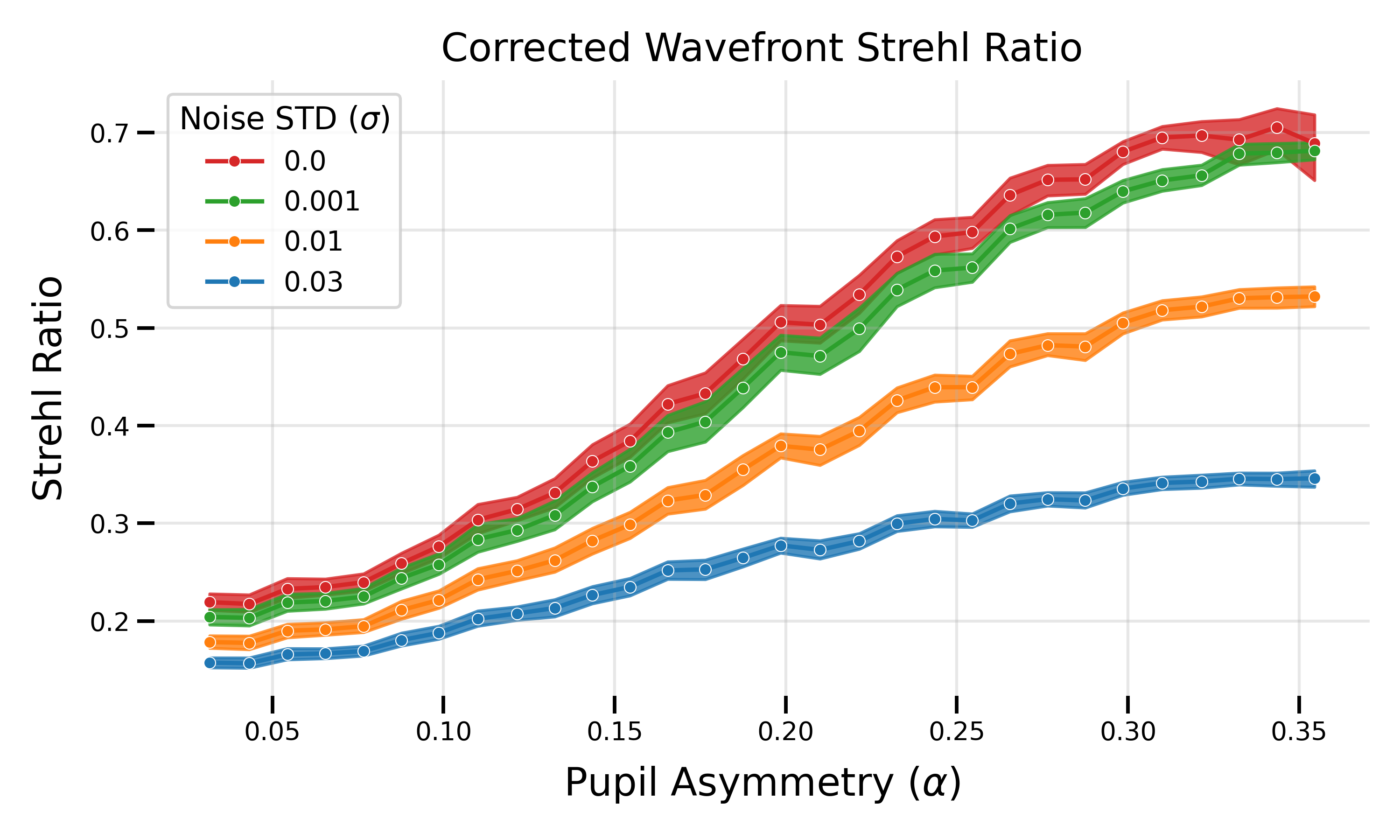} \\
        (a) Wavefront estimation error & (b) Strehl ratio
    \end{tabular}   
    
    \caption{Results from experiments with MLP architecture: (a) Wavefront estimation MSE loss as a function of pupil asymmetry (lower is better), and (b) Corrected wavefront Strehl ratio as a function of pupil asymmetry (higher is better). Results reported in this figure use testing data.}
    \label{fig:general_trend_mlp}
\end{figure}

Although our results were obtained with a uniformly distributed dataset across asymmetry, we sought to verify that the results were not due to any network or architecture bias. So we repeated the experiments with a multi layer perceptron (MLP) fully connected network in addition to the results obtained with a U-Net architecture and presented in the main paper. The results demonstrated in \fref{fig:general_trend_mlp} follow a similar trend to those presented in the main text, where more asymmetry yields lower wavefront estimation error and higher Strehl ratio, further demonstrating that the asymmetry metric relates to the wavefront recovery and is suitable for pupil design.

\section{Additional Noise Results}
We demonstrated the wavefront recovery performance across three different noise levels in addition to the noiseless case in the main text. We include additional noise cases here for more visibility. The performance of all pupils gradually degrades as the measured intensity gets noisier.

\begin{figure}[htbp]
\centering
    \begin{tabular}{cc}
    \centering
        \includegraphics[width=0.40\linewidth]{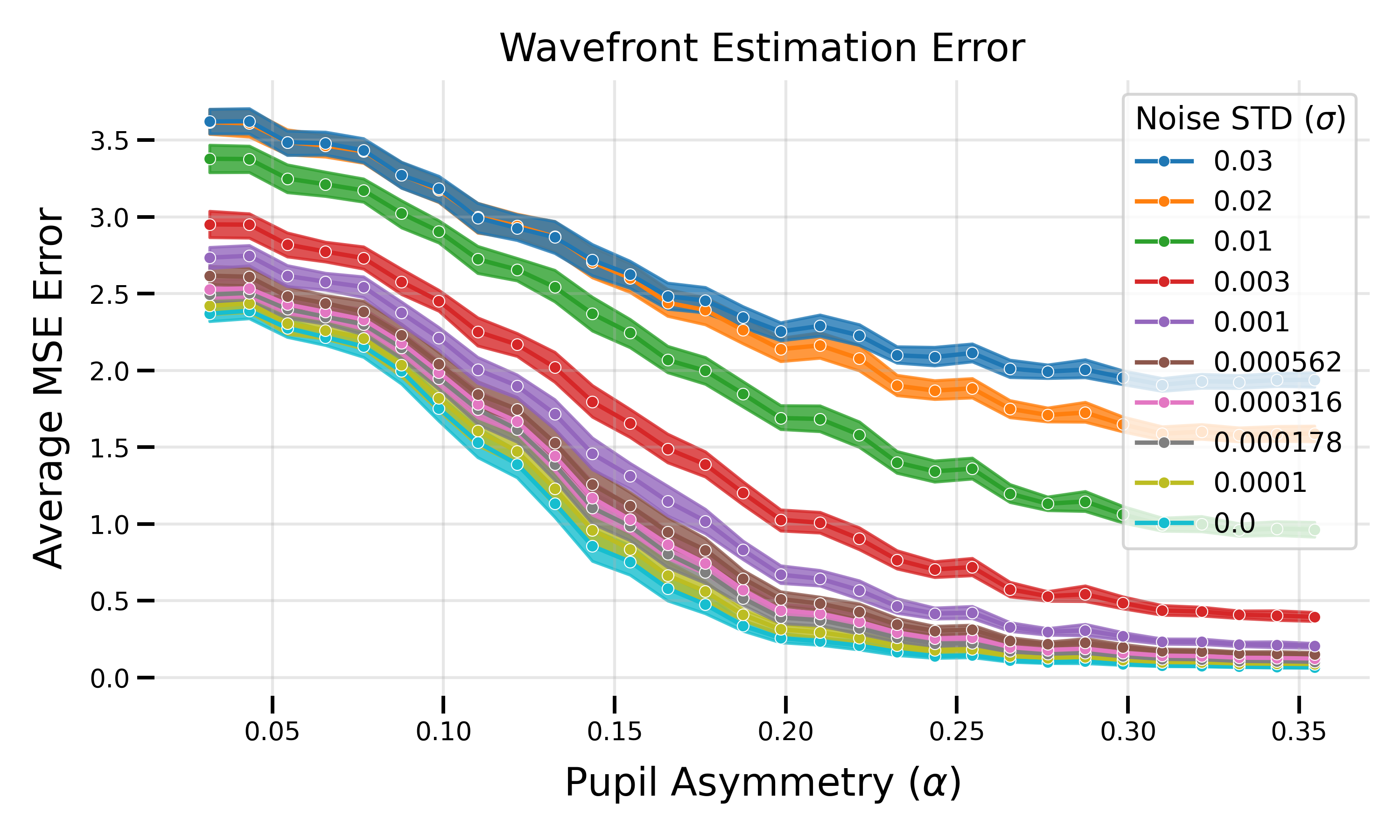} & \includegraphics[width=0.40\linewidth]{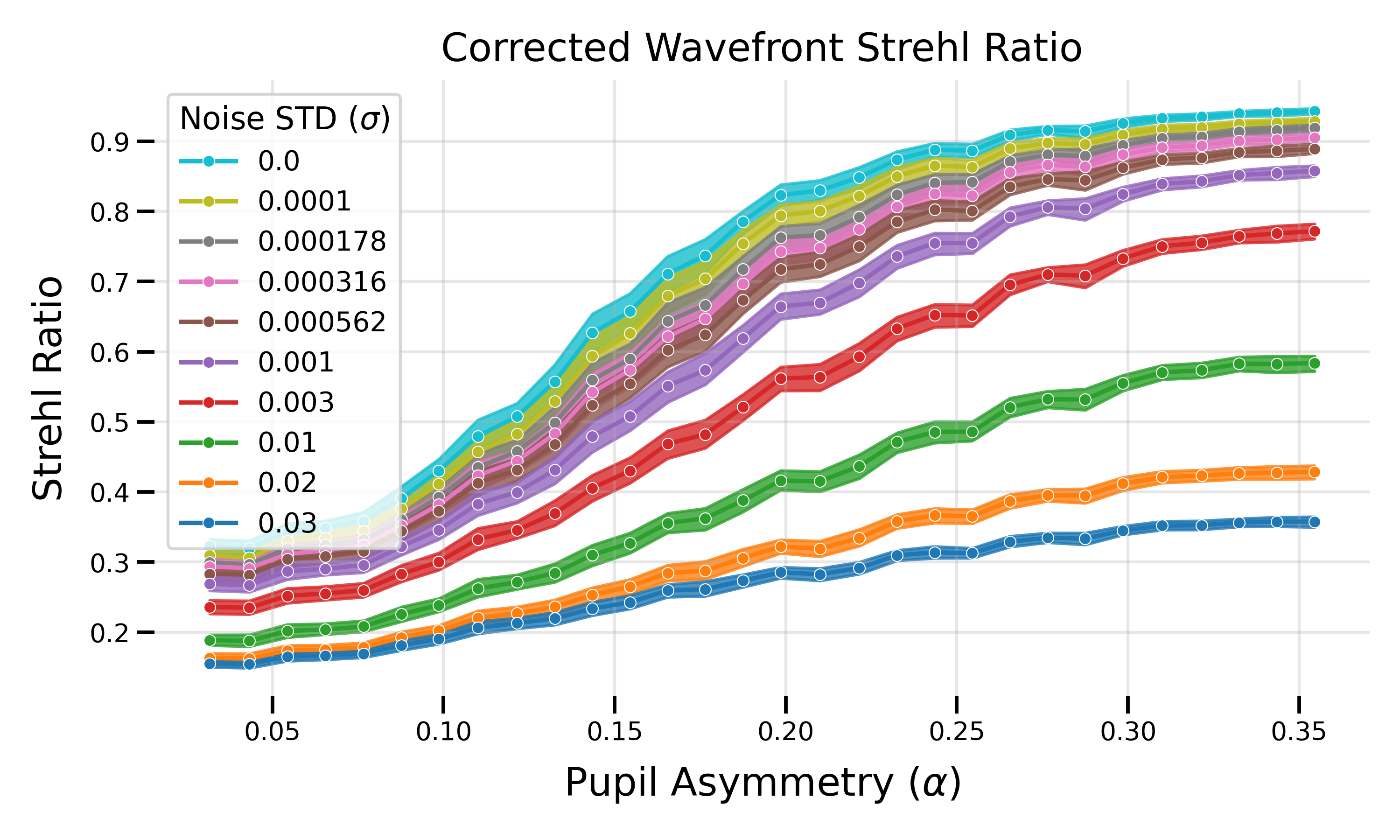} \\
        (a) Wavefront estimation error & (b) Strehl ratio
    \end{tabular}   
    
    \caption{Results under different noise conditions with the main U-Net architecture: (a) Wavefront estimation MSE loss as a function of pupil asymmetry (lower is better), and (b) Corrected wavefront Strehl ratio as a function of pupil asymmetry (higher is better). Results reported in this figure use testing data.}
    \label{fig:general_trend_more_noise_unet}
\end{figure}

\begin{figure}[htbp]
    \centering
    \includegraphics[width=0.75\linewidth]{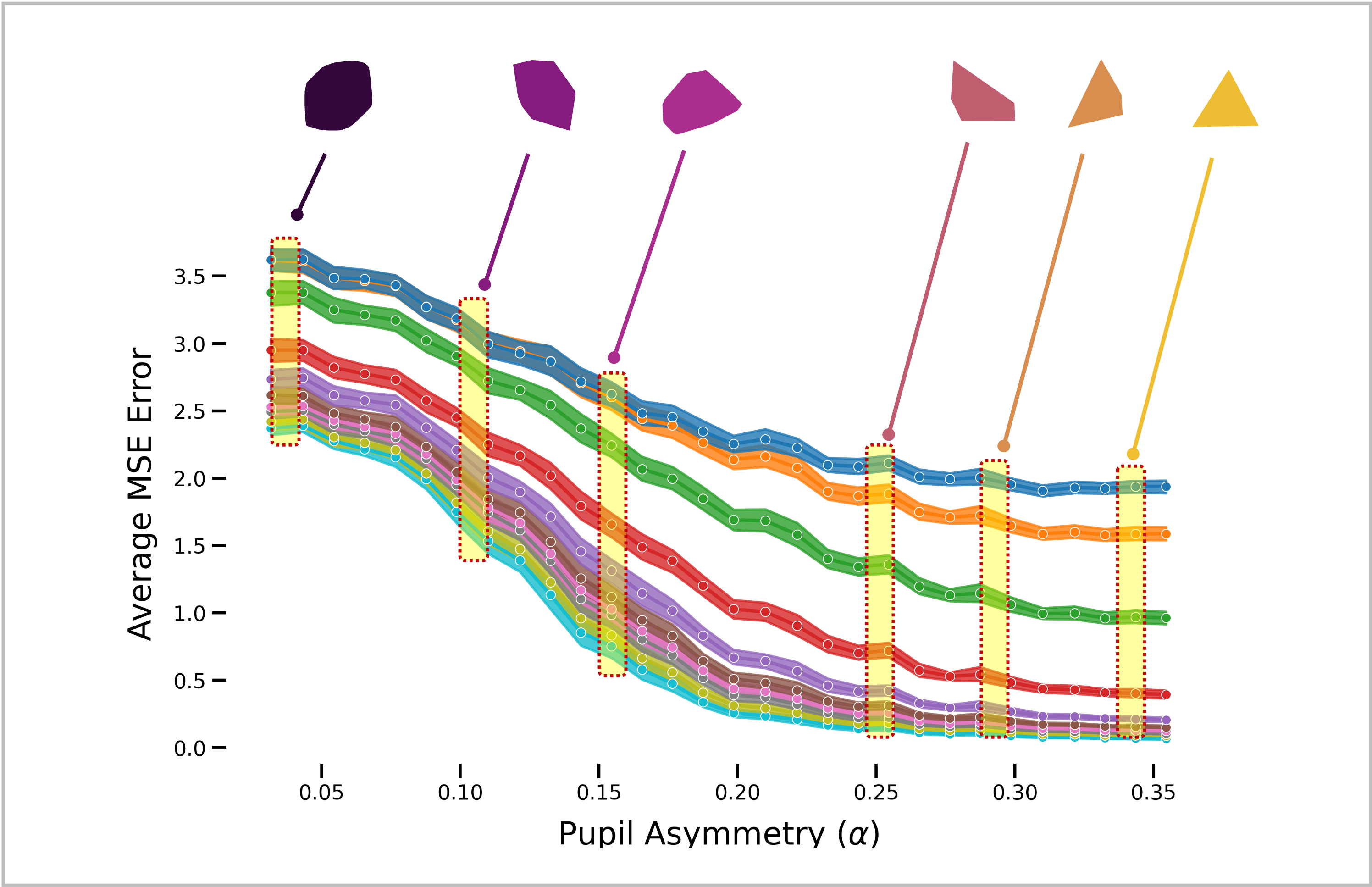}
    \caption{Visualization of how progressively more asymmetric pupils are located on the MSE loss plot.}
    \label{fig:pupils_on_loss}
\end{figure}

\section{Networks Architecture}
\label{sec:networks_arch}

\subsection{U-Net Architecture}
\begin{figure}[htbp]
    \centering
    \includegraphics[width=0.8\linewidth]{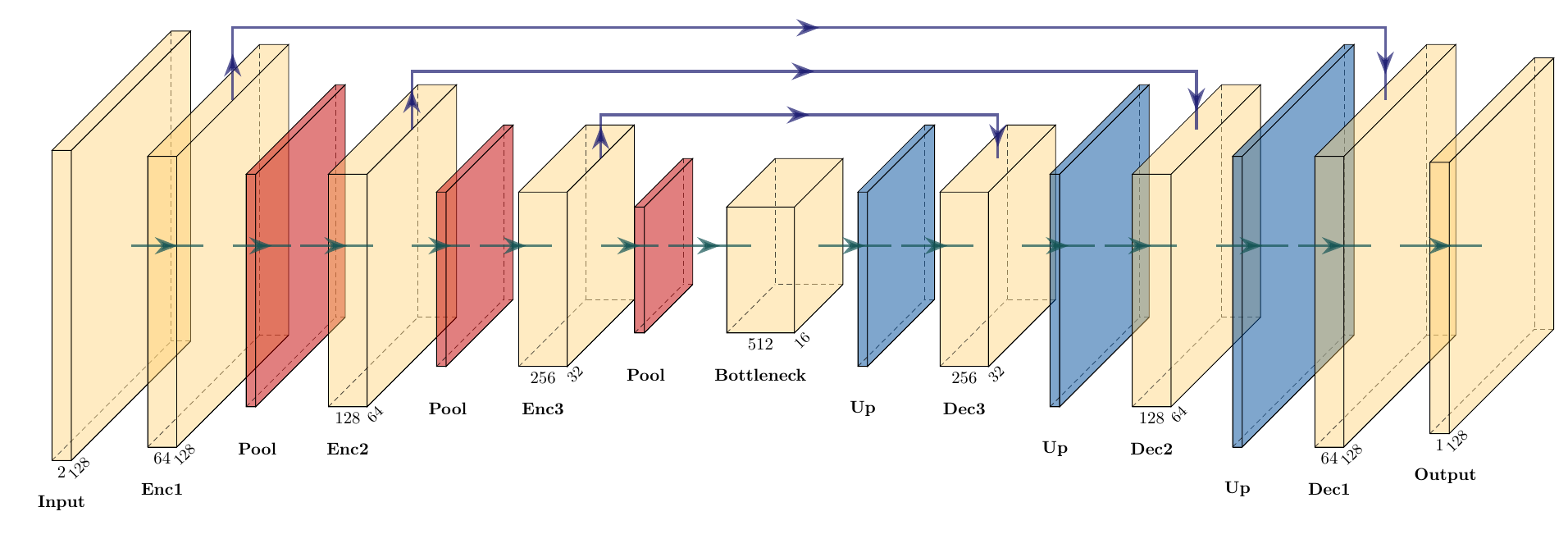}
    \caption{The architecture of the U-Net model.}
    \label{fig:unet_arch}
\end{figure}

The architecture of the U-Net model used is shown in \fref{fig:unet_arch}. The results shown in the main text are all based on this architecture.

\subsection{MLP Architecture}
\begin{figure}[htbp]
    \centering
    \includegraphics[width=0.95\linewidth]{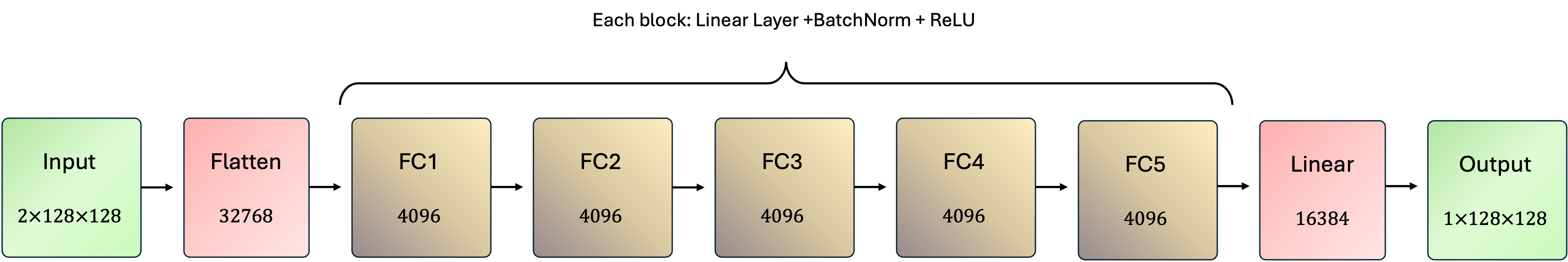}
    \caption{The architecture of the MLP model.}
    \label{fig:mlp_arch}
\end{figure}

We depict the MLP network architecture in \fref{fig:mlp_arch}. Like the U-Net model, the input to the network is the PSF and pupil mask, and the output is the phase aberrations.

\section{Additional Aberration Strength Results}
\label{sec:aberration_strength}
\begin{figure}[htbp]
    \centering
    \begin{subfigure}{0.48\linewidth}
        \centering
        \includegraphics[width=\linewidth]{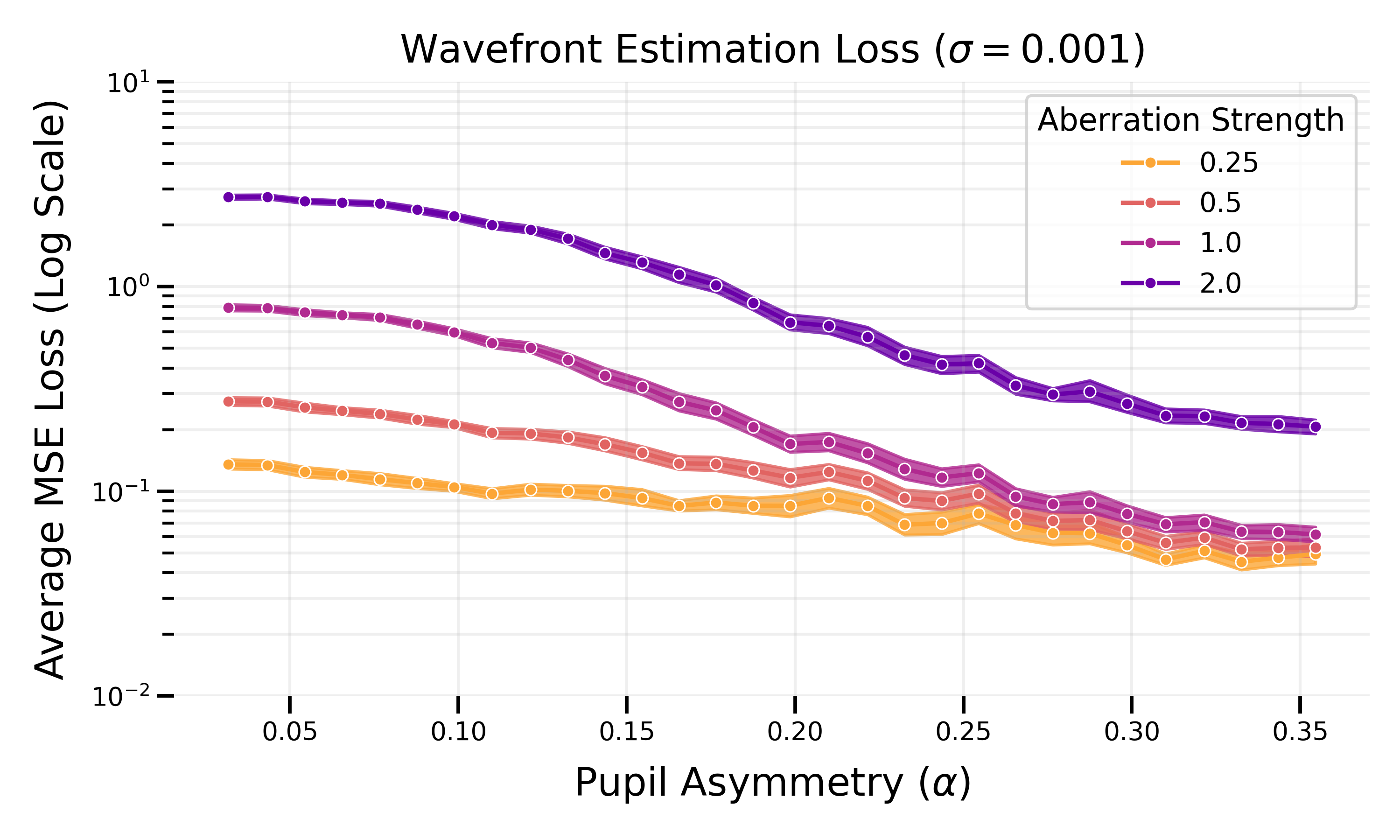}
        \caption{Wavefront Recovery MSE ($\sigma=0.001$)}
    \end{subfigure}
    \hfill
    \begin{subfigure}{0.48\linewidth}
        \centering
        \includegraphics[width=\linewidth]{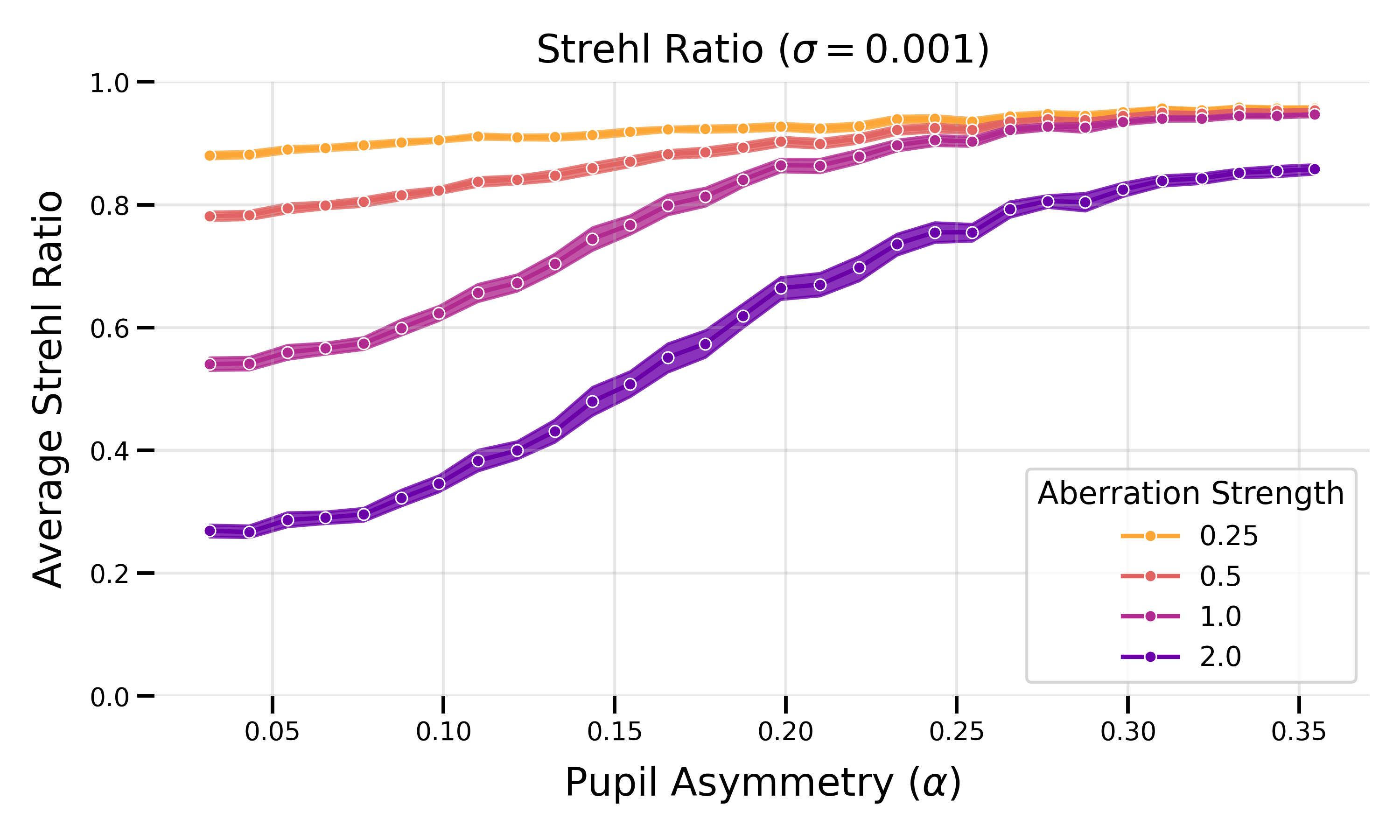}
        \caption{Strehl Ratio ($\sigma=0.001$)}
    \end{subfigure}
    \hfill
    \begin{subfigure}{0.48\linewidth}
        \centering
        \includegraphics[width=\linewidth]{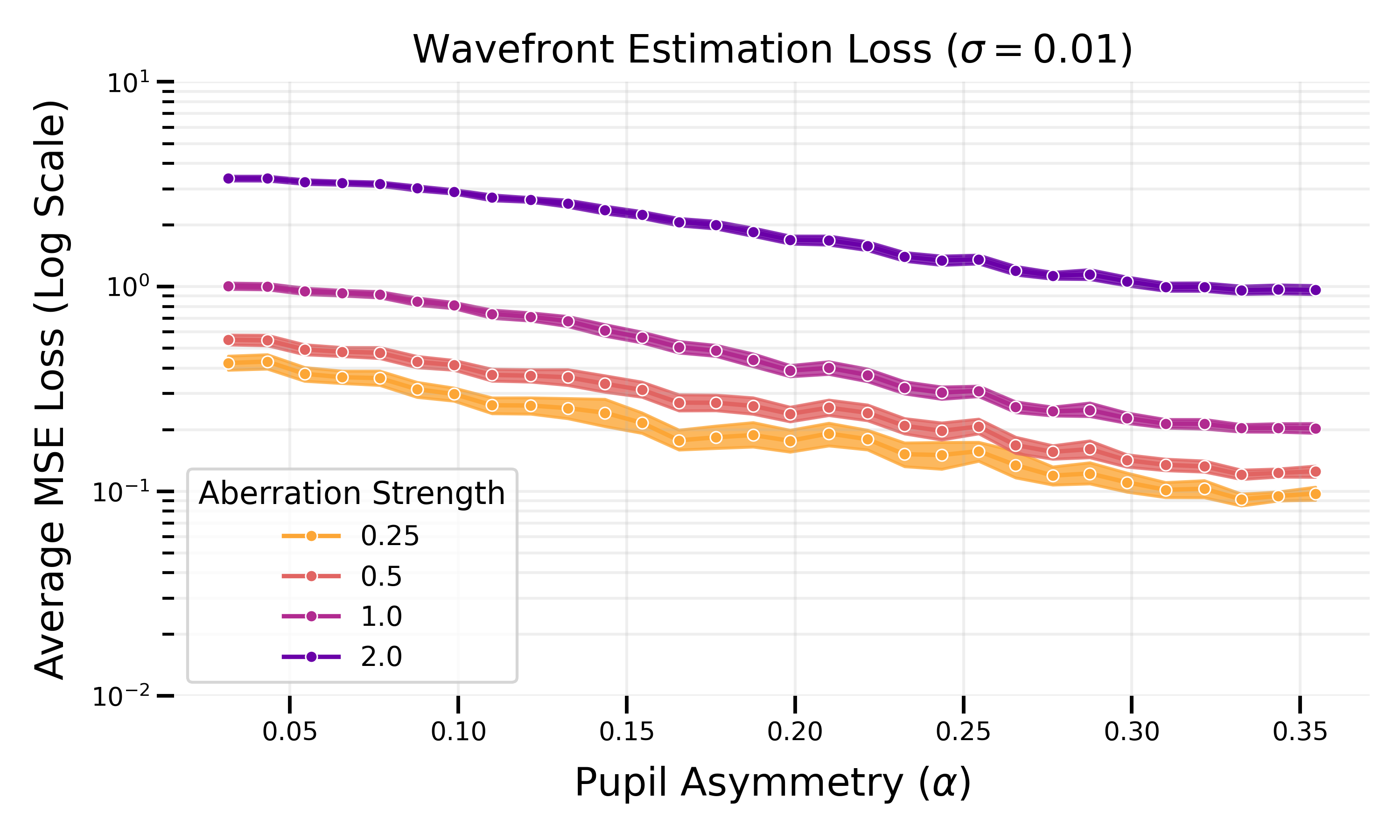}
        \caption{Wavefront Recovery MSE ($\sigma=0.01$)}
    \end{subfigure}
    \hfill
    \begin{subfigure}{0.48\linewidth}
        \centering
        \includegraphics[width=\linewidth]{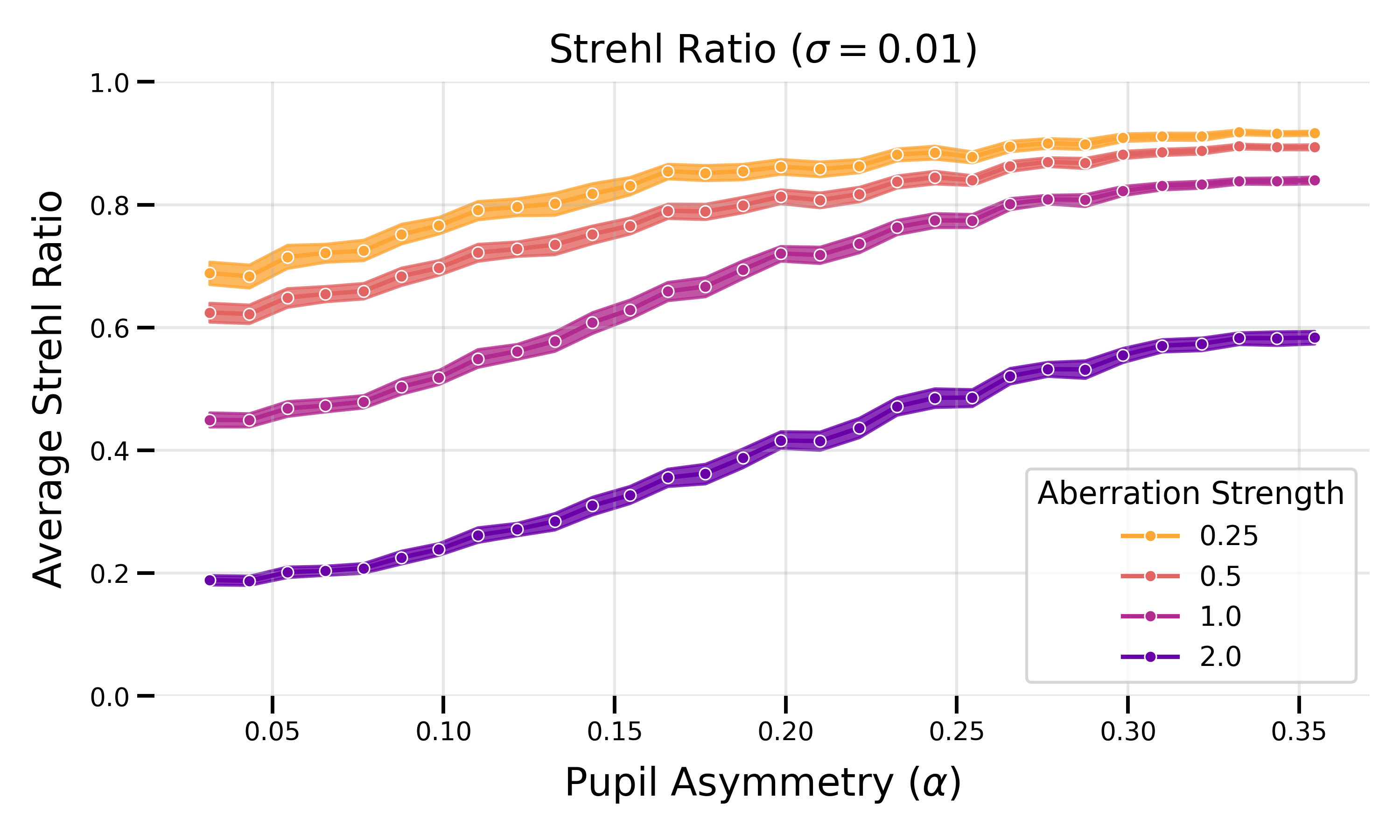}
        \caption{Strehl Ratio ($\sigma=0.01$)}
    \end{subfigure}
    \begin{subfigure}{0.48\linewidth}
        \centering
        \includegraphics[width=\linewidth]{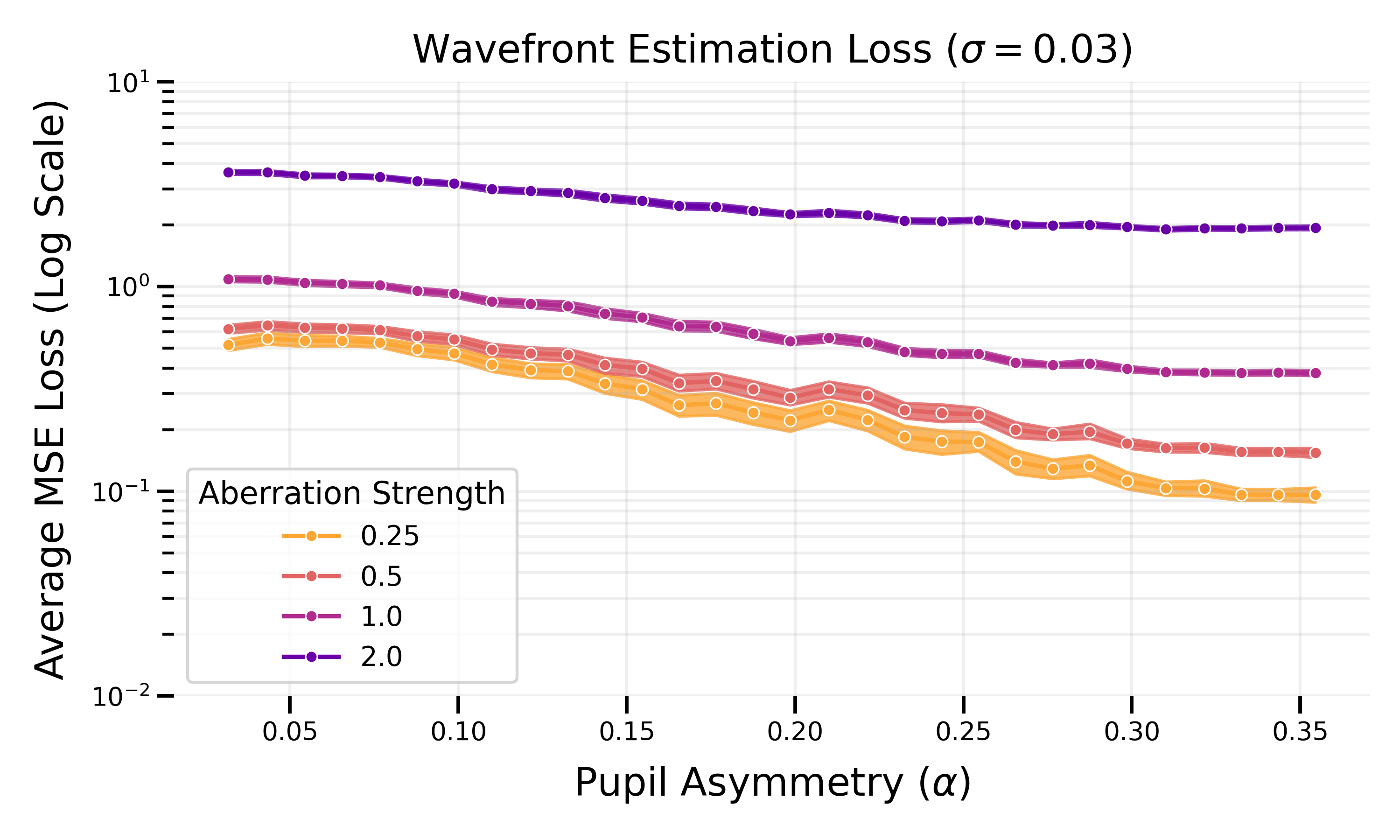}
        \caption{Wavefront Recovery MSE ($\sigma=0.03$)}
    \end{subfigure}
    \hfill
    \begin{subfigure}{0.48\linewidth}
        \centering
        \includegraphics[width=\linewidth]{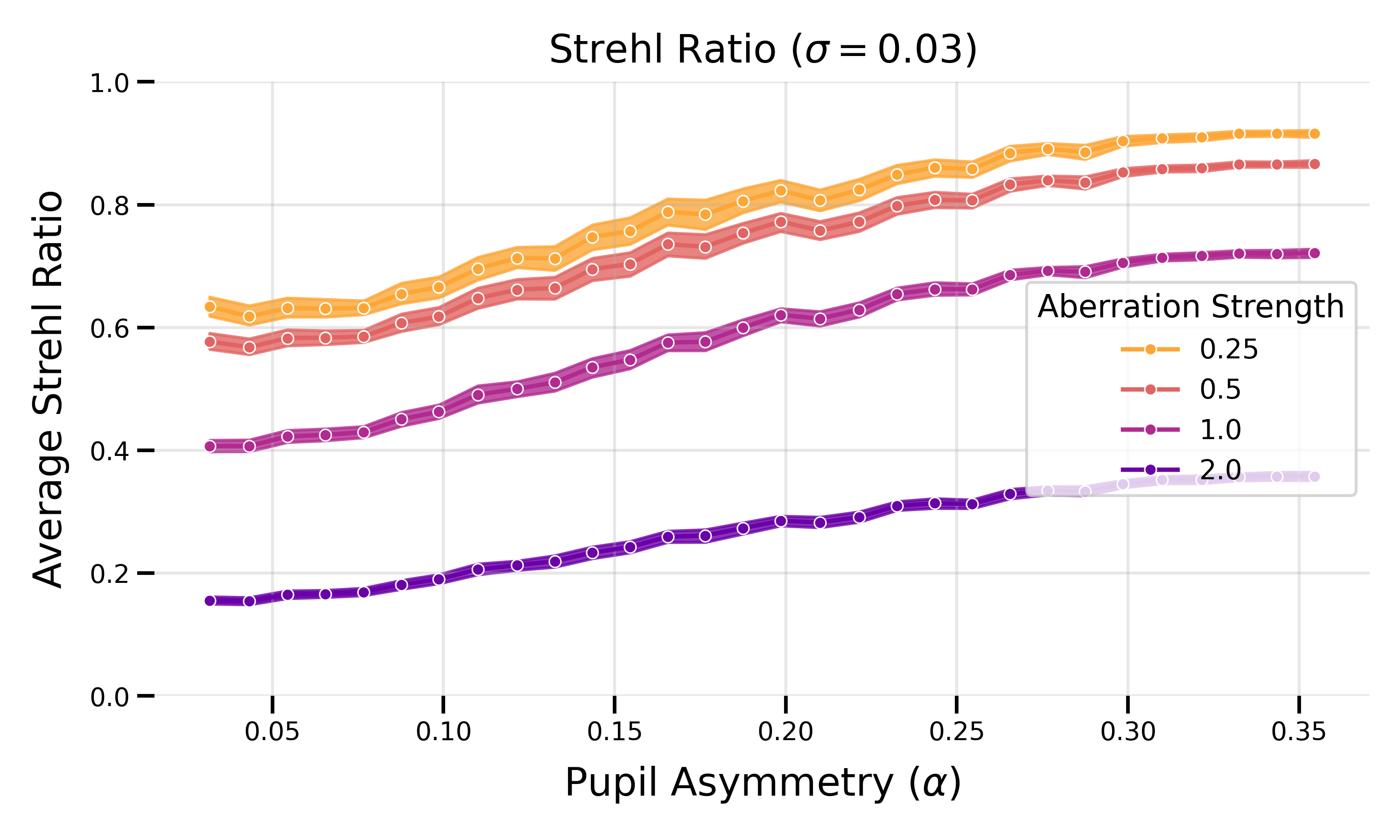}
        \caption{Strehl Ratio ($\sigma=0.03$)}
    \end{subfigure}

    \caption{Comparison of wavefront estimation performance in different aberration scenarios (0.25 to 2.0) under different noise conditions. As the aberration scale increases, symmetric pupils exhibit higher error and degradation in the Strehl ratio. The MSE is shown in log scale for visibility between different aberration strengths.}
    \label{fig:multi_experiment_scale_comparison_noisy}
\end{figure}

We reported the results of different aberration strengths under the noise-free conditions in the main text. Here, we report the results of different experiments under different noisy measurements. Like the results presented in the main text, our experiment shows that the stronger the aberrations, the bigger the performance gap between the low and high asymmetry pupils, though the gap tapers down as the measurement gets very noisy, as seen in \fref{fig:multi_experiment_scale_comparison_noisy}.

\section{Additional Real Data Prediction Results}
\label{sec:real_data_prediction}
\begin{figure}[htbp]
    \centering
    \includegraphics[width=0.95\linewidth]{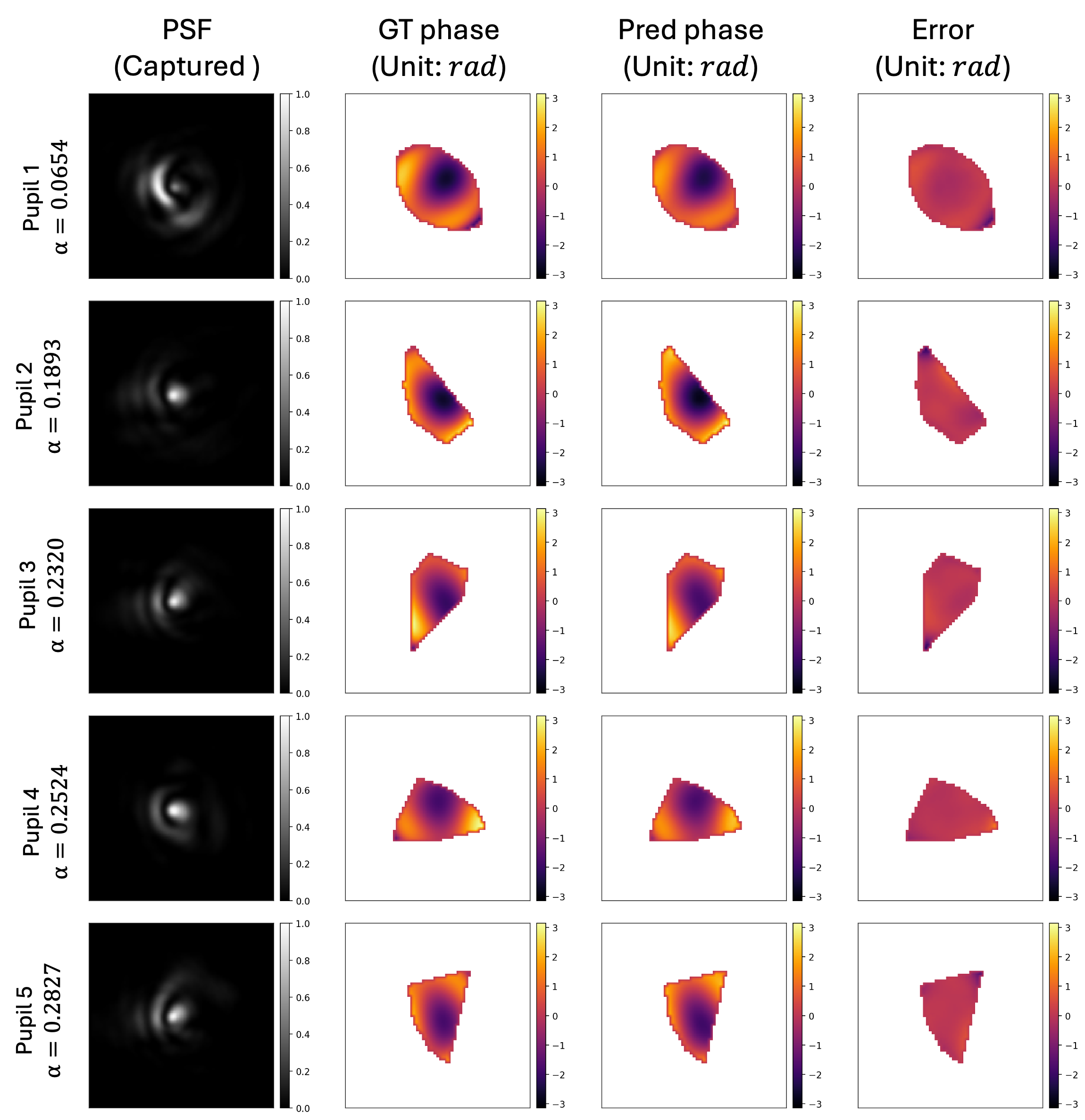}
    \caption{Wavefront estimation results from PSFs captured by our optical setup. }
    \label{fig:real_phase}
\end{figure}

\begin{figure}[htbp]
    \centering
    \includegraphics[width=0.95\linewidth]{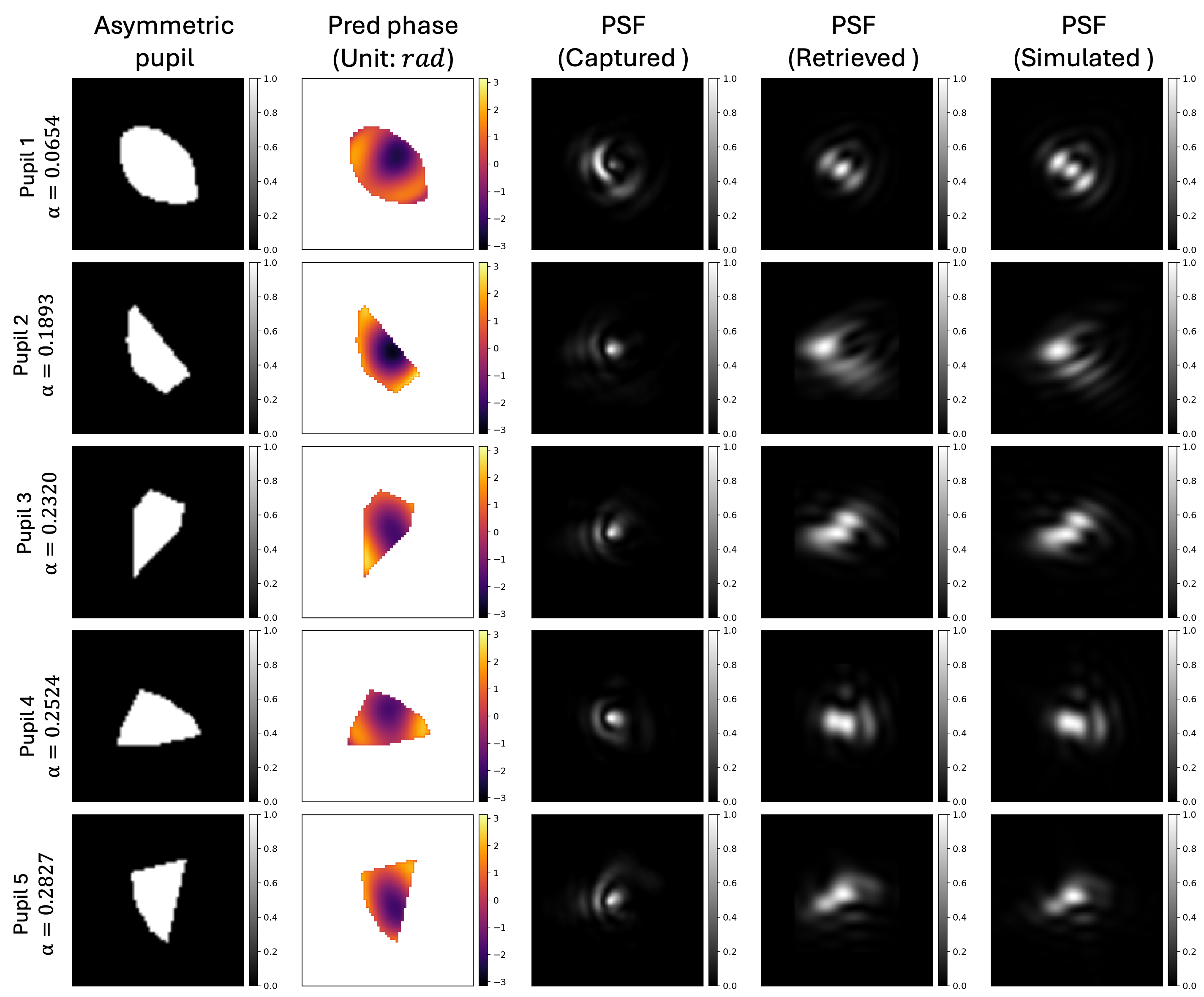}
    \caption{Visualization of PSFs constructed from the estimated wavefront.}
    \label{fig:psf_comparison}
\end{figure}

We visualize the estimated wavefront from the captured PSFs in \fref{fig:real_phase}. In Figure \ref{fig:psf_comparison}, we visualize the PSF constructed from the estimated wavefront $\vxhat$ compared with the PSF constructed from the ground truth $\vx$, which shows very close resemblance. Figure \ref{fig:psf_comparison} also demonstrates that the measured PSFs exhibit some mismatch from the intended PSF, but is sufficient in encoding the intended phase as the network generalizes to testing data. We reason this is due to two fundamental reasons: (i) our use of a checkerboard pattern for implementing the pupil, causing interference in the PSFs, and (ii) small mismatches in the optical setup. If we consider the intended PSF to be $\vy = \abs{\mF \mP \vx}^2$, where $\mP$ is the intended pupil, due to the use of the checkerboard and no explicit pupil application, the realized PSF is instead
\begin{equation}
    \widetilde{\vy} = \abs{\mF \mC \vz}^2,
\end{equation}
where $\mC$ is the larger support of the beam assumed circular. We assume that $\vz$ takes the following form:
\begin{equation}
    \vz \approx \mP \vx + (\mI - \mP) \vc,
\end{equation}
where $\mP \vx$ is the intended phase pattern, $\vc$ is the checkerboard outside of support $\mP$, and the approximation arises from small calibration errors that may exist in the pipeline.  Writing the expansion
\begin{equation}
    \widetilde{\vy} = \abs{\mF \mP \vx}^2 + \abs{\mF \mC (\mI - \mP) \vc}^2 + 2 \Re{(\mF \mP \vx) \odot (\mF \mC (\mI - \mP) \vc)^*},
\end{equation}
we can see that the checkerboard introduces an interference term that we expect to contribute to the majority of the mismatch. Since $\vc$ is constant across measurements and pupil types, the network can learn the mapping $\widetilde{\vy} \to \vx$.

\end{document}